\documentclass[aps,prd,onecolumn,groupedaddress,showpacs,nofootinbib,amssymb]{revtex4}
\usepackage[dvips]{graphicx}
\usepackage{amssymb}
\usepackage{amsmath}
\usepackage{graphicx,color}
\usepackage{amsfonts}
\usepackage{bm}
\usepackage{cancel}
\usepackage{comment}
\usepackage{hyperref}
\usepackage{ulem}

\newcommand{\GW}{\mathrm{GW}}

\newcommand{\be}{\begin{equation}}
\newcommand{\ee}{\end{equation}}
\newcommand{\bea}{\begin{eqnarray}}
\newcommand{\eea}{\end{eqnarray}}

\newcommand{\e}{\mathrm{e}}

\allowdisplaybreaks[4]

\begin{document}
\tolerance=5000

\title{Gravitational collapse, primordial black hole, and gravitational wave in Einstein--Gauss-Bonnet theory with two scalar fields}
\author{Shin'ichi~Nojiri$^{1,2}$}\email{nojiri@gravity.phys.nagoya-u.ac.jp}
\author{G.~G.~L.~Nashed$^3$}\email{nashed@bue.edu.eg}
\affiliation{$^{1)}$ Theory Center, IPNS, KEK, 1-1 Oho, Tsukuba, Ibaraki 305-0801, Japan \\
$^{2)}$ Kobayashi-Maskawa Institute for the Origin of Particles and the Universe, Nagoya University, Nagoya 464-8602, Japan \\
$^{3)}$ Center for Theoretical Physics, British University of Egypt,
Sherouk City 11837, Egypt}

\begin{abstract}
In this paper, we investigate the gravitational collapse to form the black hole in the acceleratingly expanding universe
in the frame of Einstein--Gauss-Bonnet theory having two scalar fields and we study the propagation of the gravitational wave (GW). 
This may describe the creation of the primordial black holes in the early stages of the universe and the impact of their creation on the propagation of 
primordial gravitational waves. 
The collapsing spacetime can be obtained by using the formulation of the ``reconstruction'', that is, we find a model that realises the desired or given geometry.
In the reconstructed models, ghosts often appear, which could be eliminated by imposing constraints.
We show that the standard cosmological solutions or self-gravitating objects such as a planet, the Sun, various types of stars, etc.,
in Einstein's gravity, are also solutions in this model.
Using the dynamical value of the Gauss-Bonnet coupling, the propagation of the high-frequency GW is investigated.
The propagating speed changes due to the coupling during the period of the black hole formation.
The speed of the GW propagation going into the black hole is different from that of the wave going out.
\end{abstract}

\maketitle

\section{Introduction}\label{SecI}

There are several scenarios that the recent problem of the Hubble tension could be resolved by primordial black holes (PBHs).
The PBHs are theoretical astronomical entities formed from initial disturbances in extremely dense areas through gravitational
collapse~\cite{Hawking:1971ei, Zeldovich:1967lct, Carr:1974nx, Khlopov:2008qy}.
The creation of PBH happens during the early stages of the universe, such as the time when radiation dominates or when matter equals radiation. 
The expansion history of the universe is influenced by the mass and number of PBHs~\cite{Sasaki:2016jop, Ali-Haimoud:2017rtz}.
The PBHs could be created as a result of the collapse of significant fluctuations in the initial stages of the universe.
Specifically, there are two distinct scenarios in which the collapse occurs during a radiation era and a matter era.
The presence of radiation pressure causes only significant fluctuations to collapse in the most studied case in literature.
In the second scenario, there is no pressure component involved, making it essential to accurately determine the collapse 
by considering shape deformations~\cite{Harada:2016mhb, Harada:2017fjm}.
In typical situations, there is only one period dominated by matter following the radiation-dominated era.
However, in various alternative scenarios, there could be an initial era dominated by matter before the radiation era, leading to a greater interest 
in researching the formation of PBHs during that time. 
Here, we explore the idea that PBHs could result from the collapse of primordial density fluctuations during the matter-dominated phase following a pre-big bang model.

In a spatially flat Friedmann-Lema\^{i}tre-Robertson-Walker (FLRW) universe, PBHs with masses in the range of $10^6-10^9$ $M _\mathrm{P}$ are expected 
to be formed at the end of the inflation, during the oscillatory and decay phase of the inflation field or during
the reheating~\cite{Khlopov:1980mg, Khlopov:1985fch, Garcia-Bellido:1996mdl, Cotner:2019ykd, Martin:2019nuw}.
Here $M_\mathrm{P}$ is the Planck mass.
They may also be formed naturally at high peaks of the curvature power spectrum resulting from single-field inflation~\cite{Ezquiaga:2017fvi}.
According to the scenario developed in \cite{Carr:2018nkm}, PBHs can be formed at the end of the inflationary era during a long oscillatory phase of the inflation field
that extends the reheating timescale. 
In this picture, PBHs have a characteristic mass $M\approx 4\pi \left( {M_\mathrm{P}}^2 /H_{\ast} \right)$, where $H_{\ast}$ is the Hubble scale during inflation.
In the slow-roll approximation, we have $H_{\ast}\approx (pi/16)M^2 _\mathrm{P} r A_s$ , where $r$ is the amplitude ratio between the tensor and the scalar power spectra
and $A_s$ is the amplitude of the scalar power spectrum.
Using the data from Planck-2018, one obtains $H_{\ast}\approx 10^{13}$\,GeV, which implies a typical mass of about $10^6 M_\mathrm{P}$ 
for primordial black holes formed in this scenario. 

Inflation can address the horizon, flatness, and monopole issues in hot big bang cosmology~\cite{Brout:1977ix, Guth:1980zm, Starobinsky:1980te, Linde:1981mu, Albrecht:1982wi}. 
Inflation causes quantum fluctuations to arise on tiny scales, which are then enlarged beyond the Hubble horizon and paused by the expansion 
of the inflationary cosmos~\cite{Mukhanov:1981xt}. 
Upon reentering the Hubble horizon during the decelerated expansion phase, the initial scalar perturbation acts as the foundation 
for the cosmic microwave background (CMB) irregularities and the current large-scale structure. 
This is currently estimated to be around $10^{-5}$ for scales greater than $1~\mathrm{Mpc}$~\cite{Planck:2018jri}. 
The original tensor perturbations that come back into the Hubble horizon will become the primordial gravitational waves (PGWs), 
which are seen as evidence supporting the inflationary theory. 
There are numerous experiments currently in progress or the planning stages, such as BICEP/Keck Array~\cite{BICEP:2021xfz}, 
NANOGrav~\cite{Caldwell:2022qsj}, EPTA~\cite{EPTA:2015qep}, PPTA~\cite{Zhu:2014rta}, CPTA~\cite{Xu:2023wog}, LISA~\cite{Barausse:2020rsu, LISA:2022kgy}, 
Taiji~\cite{Hu:2017mde, Ruan:2018tsw} and TianQin~\cite{TianQin:2015yph}, seeking to investigate PGW across a wide range of frequencies. 
Detecting such waves in the cosmological concordance model is challenging due to the radiation-dominated era post-inflation, 
resulting in a predicted flat GW energy spectrum $\Omega_\GW(f)\sim10^{-16} \left(r/0.01 \right)$ across $10^{-16}$ Hz to $10^8$ Hz range 
with a small tensor tilt $n_T\approx0$. 
The tensor-to-scalar ratio $r=\mathcal{P_R}/\mathcal{P}_T\lesssim0.01$~\cite{Boyle:2005se, Kuroyanagi:2008ye, Watanabe:2006qe, Saikawa:2018rcs}.

Expanding the quantum fluctuations beyond the Hubble horizon is essential for PGWs to be detectable as it leads to their classicalisation. 
The PGW has a flat energy spectrum that reaches a high frequency $f_\mathrm{UV}$, preventing the primordial tensor perturbation from being stretched beyond the horizon. 
Beyond such a high frequency $f$, a basic calculation shows a quartic ultraviolet (UV) divergence $\Omega_\GW\propto f^4$~\cite{Kuroyanagi:2008ye}, that needs to be eliminated. 
Just like how the UV divergence is handled in the energy-momentum tensor or scalar field two-point function in curved space, 
a UV regularisation method can be applied to progressively eliminate the divergence until the final regularised value is proportional to $f^{-2}$~\cite{Parker:1974qw, Fulling:1974pu}. 
The infrared spectrum could be influenced by adiabatic regularisation, resulting in unclear explanations~\cite{Durrer:2009ii, Marozzi:2011da, Markkanen:2017rvi}. 
Adiabatic regularisation is utilised in calculating the energy spectrum of PGW~\cite{Wang:2015zfa, Zhang:2018dvc}.

The identity of DM is a significant mystery in modern fundamental physics. 
An idea that has been considered for a while is that some or all dark matter could be primordial black holes~\cite{Chapline:1975ojl}, 
which are black holes formed from the collapsing of dense regions in the early universe. 
The concept that quantum fluctuations during cosmological inflation lead to the formation of overdensities responsible 
for PBH creation has been considered since the early research of \cite{Ivanov:1994pa, Garcia-Bellido:1996mdl, Bullock:1996at}. 
One specific difficulty is the requirement for fluctuations to have an amplitude approximately $10^7$ times greater than those observed 
at CMB scales ($\mathcal{P}_\zeta \sim 10^{-9}$) in order to generate a substantial number of PBHs. 
Suitable features in the single field Lagrangian or coupling the inflation to gauge fields can achieve this in models of 
single-field inflation~\cite{Kohri:2007qn, Kawasaki:2016pql, Garcia-Bellido:2017mdw, Germani:2017bcs, Motohashi:2017kbs, Ballesteros:2017fsr, 
Hertzberg:2017dkh, Cai:2018tuh, Ballesteros:2018wlw, Cai:2019bmk, Linde:2012bt, Bugaev:2013fya, Domcke:2017fix}. 
Taking into account two phases of inflation caused by separate fields provides additional opportunities for constructing 
models~\cite{Kawasaki:2016pql, Inomata:2017okj}. 
Increased variations can also result from expanding on a peak, such as in the waterfall stage 
in hybrid inflation~\cite{Garcia-Bellido:1996mdl, Lyth:2011kj, Bugaev:2011wy, Clesse:2015wea, Kawasaki:2015ppx}. 

The gravitational collapse in general relativity (GR) has been thoroughly studied since Oppenheimer, Snyder, and Datt (OSD)~\cite{Oppenheimer:1939ue},
proving that when a spherical mass collapses under its own gravity, a black hole (BH) is created.
In this basic study, the collapsing matter is represented by a uniform, pressureless fluid, i.e., dust.
When the collapse reaches its final stage, a spacetime singularity appears, which is inevitably hidden by the event horizon and invisible to distant observers.
Since then, a number of models generalising the OSD collapse scenario have been considered in academia in order to consider more realistic gravitational collapse.
According to the GR singularity theorem, under certain physical conditions (such as the energy conditions of matter), gravitational collapse inevitably results
in the formation of a spacetime singularity.
At the singularity, the density and curvature reach infinite levels~\cite{Hawking:1970zqf, Hawking:1973uf, Senovilla:2014gza}.
It is widely believed that when such an extreme spacetime singularity occurs, the classical theory of GR becomes inapplicable~\cite{Senovilla:1998oua, Malafarina:2017csn}.
Further research has shown that under physically plausible circumstances, certain collapsing matter configurations can lead to a spacetime singularity that, unlike a BH,
is not obscured by a horizon.
This suggests that spacetime singularities may be causally connected to distant observers, (see for example,~\cite{Joshi:2011rlc} and related sources).
Such events in spacetime are known as naked singularities, and some reports suggest that they can appear in modified theories of
gravity~\cite{Mosani:2021hox, Shaikh:2018cul, Abbas:2017tvh, Chatterjee:2021zre, Manna:2019tql, Ziaie:2019jfl, Dadhich:2013bya, Mkenyeleye:2014dwa}.
Although GR has successfully elucidated gravitational phenomena, the existence of spacetime singularities indicates that GR is being used outside its domain of application.
Therefore, there is a widespread consensus that GR should be replaced by an appropriately modified theory of gravity in order to tackle the singularity problem.
The emergence of spacetime singularities implies the incompleteness of paths and makes it difficult to predict the future evolution of events.
In recent decades, a great deal of effort has been devoted to identifying collapse scenarios that do not lead to a spacetime singularity.
Various studies have shown that by introducing or adjusting additional factors to the conventional modified gravity theory,
singular spacetime singularities such as those observed in GR can be avoided.
An instance demonstrating a non-singular situation can be found in $f(R)$ gravity~\cite{Bamba:2011sm}, 
where considering higher-order contributions from the Ricci scalar can provide a solution.

The generally accepted view is, however, that the final stage of the collapse process is characterised by a super-dense region of intense gravity and the emergence of Planck-scale physics.
It is believed that the non-trivial effects of quantum gravity will eventually eradicate the classical
singularity~\cite{Joshi:2008zz, Aslam:2007zz, Torres:2015aga, Modesto:2010uh, Modesto:2011kw}.
The search for a coherent model that integrates quantum effects within the context of collapse remains a central objective of many attempts in this area.
Although a comprehensive quantum theory of gravity is still incomplete, the following semi-classical approach has been used to explain the quantum aspects 
of gravity in the vicinity of singularities: near singularities, semi-classical effects can explain the quantum aspects of gravity.
This has been achieved either by quantising the space-time metric or by quantising the matter sector in the field equations of GR~\cite{Malafarina:2017csn}.
The impacts of quantum effects on the gravitational collapse of scalar fields, which typically lead to the appearance of naked singularities, have also been explored in~\cite{Husain:2008tc}.
These effects have been shown to disperse the collapsing clouds before the singularity appears.
In~\cite{Husain:2008tc}, the removal of singularities in non-uniform scalar field collapses using quantum gravity corrections is presented.
In~\cite{Bambi:2013caa, Tippett:2011hz, Bambi:2013gva}, the possibility of avoiding singularities in a homogeneous collapse process with a modified matter source
inspired by quantum gravity effects is discussed.
For other non-singular collapse scenarios and non-singular quantum cosmology models, see,
\cite{Singh:2003au, Bojowald:2008zzb, Guo:2015eho, Fathi:2016lws, Hernandez:2018rxc, Kiefer:2019bxk, Thebault:2022dmv}.
and~\cite{Casadio:2010fw, Marto:2013soa, Bojowald:2005qw, Tavakoli:2013rna, Bambi:2013gva, Malafarina:2017csn}.

Over the next decade, modern theoretical physics and cosmology will focus heavily on GW
experiments~\cite{Hild:2010id, Baker:2019nia, Smith:2019wny, Crowder:2005nr, Smith:2016jqs, Seto:2001qf, Kawamura:2020pcg, Bull:2018lat, LISACosmologyWorkingGroup:2022jok}
and cosmological microwave background radiation (CMB) and the fourth stage of the CMB~\cite{CMB-S4:2016ple, SimonsObservatory:2019qwx}.
These experiments aim to explore whether inflation occurred in the early universe.
Stochastic GW background observations of NANOGrav2023~\cite{NANOGrav:2023gor} show that a highly blue-tilted inflationary period
is required to accurately explain the signal~\cite{sunnynew}.
This places significant constraints on inflationary scenarios, even in the context of standard post-inflationary cosmological models.
Theories that lead to a slightly blue-tilted period can therefore have important implications from a phenomenological perspective.
In this situation, theories of gravity influenced by strings, like the Einstein--Gauss-Bonnet (EGB) gravity, could have a significant phenomenological impact.
For an extensive review and collection of research papers on this subject, see
\cite{Nojiri:2010wj, Nojiri:2017ncd, Hwang:2005hb, Nojiri:2006je, Nojiri:2005vv, Satoh:2007gn, Yi:2018gse, Guo:2009uk, Jiang:2013gza, Kanti:2015pda,
vandeBruck:2017voa, Kanti:1998jd, Pozdeeva:2020apf, Pozdeeva:2021iwc, Koh:2014bka, Bayarsaikhan:2020jww, Chervon:2019sey, DeLaurentis:2015fea,
Nozari:2017rta, Odintsov:2018zhw, Kawai:1998ab, Yi:2018dhl, vandeBruck:2016xvt, Kleihaus:2019rbg, Bakopoulos:2019tvc, Maeda:2011zn, Bakopoulos:2020dfg,
Ai:2020peo, Odintsov:2020xji, Oikonomou:2020sij, Odintsov:2020zkl, Odintsov:2020mkz, Venikoudis:2021irr, Kong:2021qiu, Easther:1996yd, Antoniadis:1993jc,
Antoniadis:1990uu, Kanti:1995vq, Kanti:1997br, Easson:2020mpq, Rashidi:2020wwg, Odintsov:2023aaw, Odintsov:2023lbb, Oikonomou:2022xoq, Odintsov:2023weg,
Nojiri:2023jtf, TerenteDiaz:2023kgc, Kawai:2023nqs, Kawai:2021edk, Kawai:2017kqt, Choudhury:2023kam}
and references therein.
However, the neutron star merging event of GW170817~\cite{TheLIGOScientific:2017qsa, Monitor:2017mdv, GBM:2017lvd} imposes serious constraints on the EGB theory.
It imposes important constraints on the EGB theory and implies that the velocity of the GWs ought to be similar to the speed of light in a vacuum.
The scalar coupling factor of the Gauss-Bonnet (GB) invariant, which is typically denoted as $\zeta(\phi)$ and takes the form of a function,
was largely constrained by this GW170817 event, and various scenarios were developed to construct an EGB theory that aligns with the findings of GW170817
\cite{Odintsov:2020xji, Odintsov:2020sqy, Oikonomou:2021kql, Oikonomou:2022ksx}.
The condition that the propagating speed of the gravitational wave is identical was found in \cite{Odintsov:2020sqy}
in the FLRW spacetime background but after that, in~\cite {Nojiri:2023jtf},
it was found that in the static and spherically symmetric spacetime,
the propagating speed of the gravitational wave cannot be identical to that of the light and therefore the finding of GW170817 gives a strong constraint on
the Gauss-Bonnet coupling.
In this paper, we investigate the propagation of the gravitational wave in the background that the black hole is formed in the accelerating universe. 
In order to realise the background, we use the EGB model coupled with two scalar fields $\phi$ and $\chi$.
This model could also give some indications to the formation of the primordial black hole formation in the early universe. 

Through this study, we examine the gravitational collapse and the black hole formation in the acceleratingly expanding universe as de Sitter spacetime
by using EGB theory having two scalar fields.
We also consider the effects on the propagation of the GW due to the GB coupling during the creation of the black hole.
In order to realise the gravitational collapse, we formulate the ``reconstruction'', where we try to find a model that realises the desired or given geometry.
Because the reconstructed models often include ghosts, we eliminate them by imposing constraints.
It is found that the standard cosmological solutions or self-gravitating objects such as a planet, the Sun, various types of stars, etc.,
in Einstein's gravity, are also solutions in this model.
By choosing the GB coupling properly, we consider the propagation of the GW with high frequency, because the time scale like the wavelength of
the GW is, usually, significantly lesser than the scale of the variation of the background spacetime.
It has been shown~\cite{Oikonomou:2020sij, Nojiri:2023jtf} that
the propagating speed changes due to the coupling during the period of the black hole formation.
We show that the propagation speed of the GW going into the black hole is different from that of the wave going out in general. 
By investigating the expression of the speeds, we find the conditions that the propagating speed does not exceed the light speed and does not violate the causality.

In the next section, we demonstrate that general spherical and time-varying spacetime can be rewritten in the form of the Schwarzschild-type spacetime. 
We apply the formulation to the Oppenheimer-Snyder model, which is a very well-known model of the gravitational collapse of dust into the black hole.
In Section~\ref{ads}, we apply the formulation to the de Sitter spacetime and propose a spacetime which describes the gravitational collapse and the black hole formation
in the de Sitter spacetime.
In Section~\ref{SecII}, mainly based on~\cite{Elizalde:2023rds}, we demonstrate that any spherical spacetime, regardless of its dynamism,
can be described using the EGB gravitational theory combined with the two scalar fields.
We explain how we can remove the ghosts by applying the restrictions provided by the Lagrange multiplier fields.
We show also that any solution of the Einstein equation is also a solution in this model even if we include matter.
In Section~\ref{SecIII}, we consider the GW propagation.
The final section is dedicated to summarising and discussing the findings. 

\section{Time-dependent Spherically Symmetric Spacetime}\label{Sec1}

The general expression for spherically symmetric and time-dependent spacetime is given as follows,
\begin{equation}
\label{G1}
ds^2 = - \mathcal{A}(\tau,\rho) d\tau^2 + 2 \mathcal{B} (\tau,\rho) d\tau d\rho
+ \mathcal{C}(\tau,\rho) d\rho^2 + \mathcal{D} (\tau,\rho)
\left( d\vartheta^2 + \sin^2\vartheta d\varphi^2 \right) \, .
\end{equation}
It is important to mention that the FLRW Universe with spatial flatness is a specific case within the broader class of the aforementioned spacetime.
In this section, we show that the spacetime in (\ref{G1}) can be transformed into the Schwarzschild-type spacetime as follows,
\begin{align}
\label{GBiv0}
ds^2 = - \e^{2\nu (r,t)} dt^2 + \e^{2\lambda (r,t)} dr^2 + r^2 \left( d\vartheta^2 + \sin^2\vartheta d\varphi^2 \right)\, .
\end{align}
After that, we discuss the Oppenheimer-Snyder model by applying the transformation.

\subsection{Schwarzschild-type spacetime}

For the metric given by (\ref{G1}), we redefine the radial coordinate $r$ using the following form,
\begin{equation}
\label{G2}
r^2 \equiv \mathcal{D} (\tau,\rho) \, ,
\end{equation}
 provided that $\mathcal{D} (\tau,\rho)>0$.
Generally, Eq.~(\ref{G2}) is solvable with respect to $\rho$, i.e., $\rho=\rho(\tau, r)$.
Subsequently, Eq.~(\ref{G1}) can be reformulated in the following form,
\begin{align}
\label{G3}
ds^2 =&\, \left\{ - \mathcal{A}\left(\tau,\rho\left(\tau,r\right) \right)
+ 2 \mathcal{B} \left(\tau,\rho\left(\tau,r\right) \right) \frac{\partial\rho}{\partial\tau}
+ \mathcal{C} \left(\tau,\rho\left(\tau,r\right) \right) \left( \frac{\partial\rho}{\partial\tau} \right)^2
\right\} d\tau^2 \nonumber \\
&\, + 2 \left( \mathcal{B} \left(\tau,\rho\left(\tau,r\right) \right)
+ \mathcal{C} \left(\tau,\rho\left(\tau,r\right) \right) \frac{\partial\rho}{\partial\tau} \right)
\frac{\partial \rho}{\partial r} d\tau dr \nonumber \\
&\, + \mathcal{C}\left(\tau,\rho\left(\tau,r\right) \right)
\left( \frac{\partial \rho}{\partial r} \right)^2 dr^2
+ r^2 \left( d\vartheta^2 + \sin^2\vartheta d\varphi^2 \right) \, .
\end{align}
Additionally, we represent $\tau$ as $\tau(t,r)$.
Hence, Eq. ~(\ref{G3}) can be expressed as,
\begin{align}
\label{G4}
ds^2 =&\, \left\{ - \mathcal{A}\left(\tau\left(t,r\right),
\rho\left(\tau\left(t,r\right),r\right) \right)
+ 2 \mathcal{B} \left(\tau\left(t,r\right),\rho\left(\tau,r\right)
\right) \frac{\partial\rho\left(\tau,r\right)}{\partial\tau}
+ \mathcal{C} \left(\tau,\rho\left(\tau,r\right) \right) \left( \frac{\partial\rho\left(\tau,r\right)}{\partial\tau} \right)^2
\right\} \nonumber \\
&\, \quad \times \left( \frac{\partial \tau\left(t,r\right)}{\partial t} \right)^2 dt^2 \nonumber \\
&\, + 2 \left[ \left( \mathcal{B} \left(\tau\left(t,r\right),\rho\left(\tau\left(t,r\right),r\right) \right)
+ \mathcal{C} \left(\tau,\rho\left(\tau,r\right) \right) \frac{\partial \rho\left(\tau,r\right)}{\partial \tau} \right)
\frac{\partial \rho\left(\tau,r\right)}{\partial r}
\frac{\partial \tau\left(t,r\right)}{\partial t} \right. \nonumber \\
&\, \qquad + \left\{ - \mathcal{A}\left(\tau\left(t,r\right),
\rho\left(\tau\left(t,r\right),r\right) \right)
+ 2 \mathcal{B} \left(\tau\left(t,r\right),\rho\left(\tau\left(t,r\right),r\right)
\right) \frac{\partial\rho\left(\tau,r\right)}{\partial\tau} \right. \nonumber \\
&\, \left. \left. \qquad \qquad + \mathcal{C} \left(\tau,\rho\left(\tau,r\right) \right) \left( \frac{\partial\rho\left(\tau,r\right)}{\partial\tau} \right)^2 \right\}
\frac{\partial \tau\left(t,r\right)}{\partial t}
\frac{\partial \tau\left(t,r\right)}{\partial r} \right] dt dr \nonumber \\
&\, + \left[ \mathcal{C}\left(\tau,\rho\left(\tau,r\right) \right)
\left( \frac{\partial \rho\left(\tau,r\right)}{\partial r} \right)^2
+ 2 \left( \mathcal{B} \left(\tau,\rho\left(\tau\left(t,r\right),r\right) \right)
+ \mathcal{C} \left(\tau,\rho\left(\tau\left(t,r\right),r\right) \right) \frac{\partial\rho\left(\tau,r\right)}{\partial\tau} \right)
\frac{\partial \rho\left(\tau,r\right)}{\partial r}
\frac{\partial \tau\left(t,r\right)}{\partial r} \right. \nonumber \\
&\, \qquad + \left\{ - \mathcal{A}\left(\tau\left(t,r\right),
\rho\left(\tau\left(t,r\right),r\right) \right)
+ 2 \mathcal{B} \left(\tau\left(t,r\right),\rho\left(\tau\left(t,r\right),r\right)
\right) \frac{\partial\rho\left(\tau,r\right)}{\partial\tau} \right. \nonumber \\
&\, \left. \left. \qquad \qquad + \mathcal{C} \left(\tau,\rho\left(\tau,r\right) \right) \left( \frac{\partial\rho\left(\tau,r\right)}{\partial\tau} \right)^2
\right\} \left( \frac{\partial \tau\left(t,r\right)}{\partial r} \right)^2
\right] dr^2 \nonumber \\
&\, + r^2 \left( d\vartheta^2 + \sin^2\vartheta d\varphi^2 \right) \, .
\end{align}
We can choose the time-coordinate $t$ so that,
\begin{align}
\label{G5}
0=&\, \left( \mathcal{B} \left(\tau\left(t,r\right),\rho\left(\tau\left(t,r\right),r\right) \right)
+ \mathcal{C} \left(\tau,\rho\left(\tau,r\right) \right) \frac{\partial \rho\left(\tau,r\right)}{\partial \tau} \right)
\frac{\partial \rho\left(\tau,r\right)}{\partial r}
\frac{\partial \tau\left(t,r\right)}{\partial t} \nonumber \\
&\, + \left\{ - \mathcal{A}\left(\tau\left(t,r\right),
\rho\left(\tau\left(t,r\right),r\right) \right)
+ 2 \mathcal{B} \left(\tau\left(t,r\right),\rho\left(\tau\left(t,r\right),r\right)
\right) \frac{\partial\rho\left(\tau,r\right)}{\partial\tau}
+ \mathcal{C} \left(\tau,\rho\left(\tau,r\right) \right) \left( \frac{\partial\rho\left(\tau,r\right)}{\partial\tau} \right)^2
\right\} \nonumber \\
&\, \qquad \times \frac{\partial \tau\left(t,r\right)}{\partial t}
\frac{\partial \tau\left(t,r\right)}{\partial r} \, .
\end{align}
Through the above information, $\nu(t,r)$ and $\lambda(t,r)$ can yield the following forms,
\begin{align}
\label{GBiv}
 - \e^{2\nu (r,t)} \equiv &\, \left\{ - \mathcal{A}\left(\tau\left(t,r\right), \rho\left(\tau\left(t,r\right),r\right) \right)
+ 2 \mathcal{B} \left(\tau\left(t,r\right),\rho\left(\tau\left(t,r\right),r\right) \right)
\left. \frac{\partial\rho\left(\tau,r\right)}{\partial\tau} \right|_{\tau=\tau\left(t,r\right)} \right. \nonumber \\
&\, \left. \qquad + \mathcal{C} \left( \tau,\rho\left(\tau,r\right) \right) \left( \left. \frac{\partial\rho\left(\tau,r\right)}{\partial\tau} \right|_{\tau=\tau\left(t,r\right)} \right)^2
\right\} \left( \frac{\partial \tau\left(t,r\right)}{\partial t} \right)^2 \, , \nonumber \\
\e^{2\lambda (r,t)} \equiv\, & \mathcal{C}\left(\tau,\rho\left(\tau,r\right) \right)
\left( \left. \frac{\partial \rho\left(\tau,r\right)}{\partial r} \right|_{\tau=\tau\left(t,r\right)} \right)^2 \nonumber \\
&\, + 2 \left( \mathcal{B} \left(\tau,\rho\left(\tau\left(t,r\right),r\right) \right)
+ \mathcal{C} \left(\tau,\rho\left(\tau\left(t,r\right),r\right) \right) \left. \frac{\partial\rho\left(\tau,r\right)}{\partial\tau}\right|_{\tau=\tau\left(t,r\right)} \right)
\left. \frac{\partial \rho\left(\tau,r\right)}{\partial r} \right|_{\tau=\tau\left(t,r\right)} \frac{\partial \tau\left(t,r\right)}{\partial r} \nonumber \\
&\, + \left\{ - \mathcal{A}\left(\tau\left(t,r\right), \rho\left(\tau\left(t,r\right),r\right) \right)
+ 2 \mathcal{B} \left(\tau\left(t,r\right),\rho\left(\tau\left(t,r\right),r\right)
\right) \left. \frac{\partial\rho\left(\tau,r\right)}{\partial\tau} \right|_{\tau=\tau\left(t,r\right)} \right. \nonumber \\
&\, \left. \qquad \qquad + \mathcal{C} \left(\tau,\rho\left(\tau,r\right) \right) \left( \left. \frac{\partial\rho\left(\tau,r\right)}{\partial\tau} \right|_{\tau=\tau\left(t,r\right)} \right)^2
\right\} \left( \frac{\partial \tau\left(t,r\right)}{\partial r} \right)^2 \, .
\end{align}
Thus, we have demonstrated that the most comprehensive form of the spherically symmetric and time-dependent spacetime in (\ref{G1}) can be rewritten in the form of
the Schwarzschild-type spacetime in (\ref{GBiv0}).
Thus, we can begin with Eq.~(\ref{GBiv0}) when addressing the spherically symmetric and time-dependent spacetime without losing any generality

\subsection{Oppenheimer-Snyder model}

The Oppenheimer-Snyder model is the simplest model describing the collapse of a star into a black hole.
The model is given by the collapsing ball of dust.
Inside the ball, the spacetime is described by the shrinking FLRW spacetime with dust.
Although the Oppenheimer-Snyder model is very well known, some details have not been so clear.
We now briefly review the model in this subsection.

For the FLRW universe, we find
\begin{align}
\label{FLRW_ABCD}
\mathcal{A}=1\, , \quad \mathcal{B}=0 \, , \quad \mathcal{C}=a(\tau)^2 \, , \quad
r^2 = \mathcal{D} = a(\tau)^2 \rho^2 \, .
\end{align}
Then Eq.~(\ref{G5}) reduces
\begin{align}
\label{G5B}
0 = - \frac{a'(\tau) r}{a(\tau)} + \left( -1 + \frac{a'(\tau)^2 r^2}{a(\tau)^2} \right)
\frac{\partial \tau\left(t,r\right)}{\partial r} \, .
\end{align}
Here we have used $\rho=\frac{r}{a(\tau)}$, which is given by the last equation in (\ref{FLRW_ABCD}).
Eq.~(\ref{GBiv}) also gives
\begin{align}
\label{GBivBB}
 - \e^{2\nu (r,t)} =&\, \left( -1 + \frac{a'(\tau)^2 r^2}{a(\tau)^2} \right)
\left( \frac{\partial \tau\left(t,r\right)}{\partial t} \right)^2 \, , \nonumber \\
\e^{2\lambda (r,t)} =&\, 1 { - \frac{2r a'(\tau)}{a(\tau)} \frac{\partial \tau\left(t,r\right)}{\partial r} }
+ \left( -1 + \frac{a'(\tau)^2 r^2}{a(\tau)^2} \right) \left( \frac{\partial \tau\left(t,r\right)}{\partial r} \right)^2 \, .
\end{align}
We now consider the shrinking dust ball, where $a= a_0 \left( - \tau \right)^\frac{2}{3}$ with a constant $a_0$ and the surface of the ball is given by $\rho=\rho_0$.
We assume $\tau$ is negative and the ball shrinks to a point at $\tau=0$.
Then the equations in (\ref{GBivBB}) have the following froms
\begin{align}
\label{GBivBB2}
 - \e^{2\nu (r,t)} = \left( -1 + \frac{4 r^2}{9 \tau^2} \right)
\left( \frac{\partial \tau\left(t,r\right)}{\partial t} \right)^2 \, , \quad
\e^{2\lambda (r,t)} = 1 { - \frac{4r}{3\tau} \frac{\partial \tau\left(t,r\right)}{\partial r} }
+ \left( -1 + \frac{4 r^2}{9 \tau^2} \right) \left( \frac{\partial \tau\left(t,r\right)}{\partial r} \right)^2 \, ,
\end{align}
and Eq.~(\ref{G5B}) is rewritten as
\begin{align}
\label{G5B2}
0 = - \frac{2r}{3\tau}
+ \left( -1 + \frac{4 r^2}{9 \tau^2} \right)
\frac{\partial \tau\left(t,r\right)}{\partial r} \, .
\end{align}
If we define $\mathcal{T}(t,r)\equiv \frac{\tau(t,r)}{r}$, Eq.~(\ref{G5B2}) has the following form,
\begin{align}
\label{G5B3}
0 = - \frac{2}{3\mathcal{T}} + \left( -1 + \frac{4}{9 \mathcal{T}^2} \right)
\left( \mathcal{T} + r \frac{\partial \mathcal{T}}{\partial r} \right) \, ,
\end{align}
whose general solution is given by
\begin{align}
\label{G5B5}
\frac{r}{r_0(t)} = {\mathcal{T}}^2 \left({\mathcal{T}}^2 + \frac{2}{9} \right)^{-\frac{3}{2}} \, .
\end{align}
Here $r_0(t)$ is an arbitrary function of $t$.
By using the definition of $\mathcal{T}(t,r)$, $\mathcal{T}(t,r)\equiv \frac{\tau(t,r)}{r}$, we can further rewrite (\ref{G5B5})
as follows
\begin{align}
\label{G5B5B}
\frac{\tau(t,r)}{r_0(t)} = - \left(1 + \frac{2r^2}{9\tau(t,r)^2} \right)^{-\frac{3}{2}} \, ,
\end{align}
which gives,
\begin{align}
\label{G5B6}
0= \left\{ - \frac{1}{r_0(t)} - \frac{2}{3} \left(1 + \frac{2r^2}{9\tau(t,r)^2} \right)^{-\frac{5}{2}} \frac{r^2}{\tau(t,r)^3}\right\}
\frac{\partial \tau\left(t,r\right)}{\partial r}
+ \frac{2}{3} \left(1 + \frac{2r^2}{9\tau(t,r)^2} \right)^{-\frac{5}{2}} \frac{r}{\tau(t,r)^2} \, .
\end{align}
Furthermore, we can rewrite (\ref{G5B6}) as in (\ref{G5B2}),
\begin{align}
\label{G5B7}
\frac{\partial \tau\left(t,r\right)}{\partial r}
= - \frac{\frac{2r}{3\tau}}{1 - \frac{4r^2}{9\tau^2}} \, .
\end{align}
We also find
\begin{align}
\label{dtaudt1}
- \frac{\tau(t,r) r_0'(t) }{r_0(t)^2} = \left\{ - \frac{1}{r_0(t)} - \frac{2}{3} \left(1 + \frac{2r^2}{9\tau(t,r)^2} \right)^{-\frac{5}{2}} \frac{r^2}{\tau(t,r)^3}\right\}
\frac{\partial \tau\left(t,r\right)}{\partial t} \, ,
\end{align}
and therefore
\begin{align}
\label{dtaudt2}
\frac{\partial \tau\left(t,r\right)}{\partial t} = - \frac{\left(1 + \frac{2r^2}{9\tau(t,r)^2} \right)^{-\frac{1}{2}} r_0'(t)}{1 - \frac{4r^2}{9\tau^2}} \, .
\end{align}
By combining (\ref{GBivBB2}), (\ref{G5B2}), (\ref{G5B5}), and (\ref{G5B7}), we find
\begin{align}
\label{GBivBB3}
 - \e^{2\nu (r,t)} = - \frac{\left(1 + \frac{2r^2}{9\tau(t,r)^2} \right)^{-1} r_0'(t)^2}{1 - \frac{4 r^2}{9 \tau^2} } \, , \quad
\e^{2\lambda (r,t)} = \frac{1}{1 - \frac{4r^2}{9\tau^2}} \, .
\end{align}

We are now considering the shrinking dust ball, whose surface exists at $\rho=\rho_0$.
Outside the ball, $\rho>\rho_0$, we assume the Schwarzschild metric as
\begin{align}
\label{Schwarzschild}
ds^2 = - \left( 1 - \frac{2M}{r} \right) dt^2 + \left( 1 - \frac{2M}{r} \right)^{-1} dr^2 + r^2 \left( d\vartheta^2 + \sin^2\vartheta d\varphi^2 \right)\, .
\end{align}
Then this model is nothing but the Oppenheimer-Snyder model~\cite{Oppenheimer:1939ue}.
We should note that the fluid besides the dust cannot make a ball with a uniform density due to the pressure.
At $\rho=\rho_0$, that is, $r=a\left( \tau \right) \rho_0 = \rho_0 a_0 \left( - \tau \right)^\frac{2}{3}$, we find
\begin{align}
\label{GBivBB3BB1}
 - \e^{2\nu} = - \frac{\left(1 + \frac{2{\rho_0}^2 {a_0}^2}{9\left( -\tau\right)^\frac{2}{3}} \right)^{-1} r_0'(t)^2}{1 - \frac{4 {\rho_0}^2 {a_0}^2}{9\left( -\tau\right)^\frac{2}{3}} } \, , \quad
\e^{2\lambda} = \frac{1}{1 - \frac{4{\rho_0}^2 {a_0}^2}{9\left( -\tau\right)^\frac{2}{3}}} \, .
\end{align}
We now choose $r_0(t)$ so that
\begin{align}
\label{r0}
r_0'(t)^2 = \left(1 + \frac{2{\rho_0}^2 {a_0}^2}{9\left( -\tau\right)^\frac{2}{3}} \right) \left( 1 - \frac{4 {\rho_0}^2 {a_0}^2}{9\left( -\tau\right)^\frac{2}{3}} \right)^2 \, ,
\end{align}
on the surface of the dust ball.
Then we find
\begin{align}
\label{surface1}
\e^{2\nu}=\e^{-2\lambda}= 1 - \frac{4 {\rho_0}^3 {a_0}^3}{9r}\, ,
\end{align}
on the surface.
By comparing (\ref{surface1}) with (\ref{Schwarzschild}), we find
\begin{align}
\label{mass}
M = \frac{2 {\rho_0}^3 {a_0}^3}{9} \, .
\end{align}
On the surface, Eq.~(\ref{G5B5}) has the following form,
\begin{align}
\label{G5B5CCC}
\frac{\tau(t,r)}{r_0(t)} = - \left(1 + \frac{2{\rho_0}^2 {a_0}^2 }{9\left( - \tau \right)^\frac{2}{3}} \right)^{-\frac{3}{2}} \, ,
\end{align}
that is,
\begin{align}
\label{taur0B1}
1 + \frac{2{\rho_0}^2 {a_0}^2 }{9\left( - \tau \right)^\frac{2}{3}} = \left( - \frac{\tau}{r_0(t)} \right)^{-\frac{2}{3}}
= \frac{r_0(t)^{\frac{2}{3}}}{\left( - \tau\right)^\frac{2}{3}}\, ,
\end{align}
which gives,
\begin{align}
\label{taur0B2}
\frac{2{\rho_0}^2 {a_0}^2 }{9\left( - \tau \right)^\frac{2}{3}} = \frac{1}{\frac{9r_0(t)^\frac{2}{3}}{2{\rho_0}^2{a_0}^2} - 1}\, .
\end{align}
By substituting the above expression (\ref{taur0B2}) into (\ref{r0}), we obtain the differential equation for $r_0(t)$ as follows,
\begin{align}
\label{r0B}
r_0'(t)^2 = \frac{\frac{9r_0(t)^\frac{2}{3}}{2{\rho_0}^2{a_0}^2}\left( \frac{9r_0(t)^\frac{2}{3}}{2{\rho_0}^2{a_0}^2} - 3\right)^2}
{\left(\frac{9r_0(t)^\frac{2}{3}}{2{\rho_0}^2{a_0}^2} - 1\right)^3} \, ,
\end{align}
Therefore the formal solution of (\ref{r0B}) is given by
\begin{align}
\label{r0B2}
t = \pm \int^{r_0(t)} dq \sqrt{\frac{\left(\frac{9q^\frac{2}{3}}{2{\rho_0}^2{a_0}^2} - 1\right)^3}
{\frac{9q^\frac{2}{3}}{2{\rho_0}^2{a_0}^2}\left( \frac{9q^\frac{2}{3}}{2{\rho_0}^2{a_0}^2} - 3\right)^2}} \, .
\end{align}
Because the l.h.s. of (\ref{r0B}) is not negative, there is a lower bound
\begin{align}
\label{r0B3}
r_0(t) \geq \left(\frac{2{\rho_0}^2{a_0}^2}{9}\right)^\frac{3}{2}\, .
\end{align}
The r.h.s. of (\ref{r0B}) diverges when $r_0(t)$ goes to the lower bound.
The r.h.s. of (\ref{r0B}) also vanishes at $r_0(t) = \left(\frac{2{\rho_0}^2{a_0}^2}{3}\right)^\frac{3}{2}$.
We should further note that the r.h.s. of (\ref{r0B}) goes to unity when $r_0(t)$ goes to infinity.

When $r_0(t) \sim \left(\frac{2{\rho_0}^2{a_0}^2}{9}\right)^\frac{3}{2}$, Eq.~(\ref{r0B}) behaves as
\begin{align}
\label{r0B4}
r_0'(t)^2 \sim \frac{27 \left(\frac{2{\rho_0}^2{a_0}^2}{9}\right)^\frac{9}{2}}
{2\left( r_0(t) - \left(\frac{2{\rho_0}^2{a_0}^2}{9}\right)^\frac{3}{2} \right)^3} \, ,
\end{align}
whose solution is
\begin{align}
\label{r0B5}
\pm \frac{\frac{2}{5}\sqrt{\frac{2}{27}}\left( r_0(t) - \left(\frac{2{\rho_0}^2{a_0}^2}{9}\right)^\frac{3}{2} \right)^\frac{5}{2}}{\left(\frac{2{\rho_0}^2{a_0}^2}{9}\right)^\frac{9}{4}}
\sim t - C_t \, .
\end{align}
Here $C_t$ is a constant of the integration.
Then we obtain
\begin{align}
\label{r0B6}
r_0(t) \sim \left(\frac{2{\rho_0}^2{a_0}^2}{9}\right)^\frac{3}{2} + \left( \frac{675}{8} \right)^\frac{1}{5} \left(\frac{2{\rho_0}^2{a_0}^2}{9}\right)^\frac{9}{10}
\left| t - C_t \right|^\frac{2}{5} \, .
\end{align}
When $t-C_t$ vanishes, there is a singularity where $r_0'(t)$ diverges as required.
Eq.~(\ref{taur0B2}) tells that when $r_0(t) \to \left(\frac{2{\rho_0}^2{a_0}^2}{9}\right)^\frac{3}{2}$, $\tau$ goes to vanish on the surface of the dust ball, which tells that
the radius of the ball also vanishes, that is, the dust ball collapses to a point as in the Big Crunch.

When $r_0(t) \sim \left(\frac{2{\rho_0}^2{a_0}^2}{3}\right)^\frac{3}{2}$, Eq.~(\ref{r0B}) tells
\begin{align}
\label{r0B7}
r_0'(t)^2 \sim \frac{3\ \left( r_0(t) - \left(\frac{2{\rho_0}^2{a_0}^2}{3}\right)^\frac{3}{2} \right)^2}{2 \left(\frac{2{\rho_0}^2{a_0}^2}{3}\right)^3} \, ,
\end{align}
whose solution is
\begin{align}
\label{r0B8}
r_0(t) \sim \left(\frac{2{\rho_0}^2{a_0}^2}{3}\right)^\frac{3}{2} + C_r \exp \left( \pm \frac{3t}{2 \left(\frac{2{\rho_0}^2{a_0}^2}{3}\right)^3} \right)\, .
\end{align}
This tells that $r_0(t) = \left(\frac{2{\rho_0}^2{a_0}^2}{3}\right)^\frac{3}{2}$ corresponds to the infinite future, which corresponds to $-$ signature in $\pm$ in the exponential function,
or infinite past, which corresponds to $+$ signature.
As we find the arguments so far, when $r_0(t) = \left(\frac{2{\rho_0}^2{a_0}^2}{3}\right)^\frac{3}{2}$, the radius of the dust ball coincides
with the Schwarzschild radius, that is, the appearance of the horizon.
In fact, on the surface of the dust ball, we find $\e^{2\nu} = \e^{-2\lambda} = 0$.
Here we used (\ref{GBivBB3BB1}) with (\ref{r0}), and (\ref{taur0B1}).
As well known, it takes infinite time if we use the time coordinate $t$, which is the time of the observer in the infinitely far from the black hole.
Because we are considering the collapse of the dust ball, we should choose $-$ in $\pm$ of Eq.~(\ref{r0B8}).

Finally when $r_0(t)\to \infty$, we find $r_0'(t)^2 \sim 1$, whose solution is
\begin{align}
\label{r0B9}
r_0(t) \sim \pm \left( t - C_t \right) \, ,
\end{align}
with a constant of the integration $C_t$.
Because $r_0(t)$ is positive, $+$ signature in $\pm$ corresponds to the infinite future and $-$ signature to the infinite past.
Eq.~(\ref{taur0B2}) tells that when $r_0(t)\to \infty$ corresponds to $-\tau\to \infty$, that is, the radius of the dust ball is infinite.
Because we are considering the collapse of the dust ball, we should choose $-$ in $\pm$ of Eq.~(\ref{r0B9}).

In summary, when we consider the collapse, we may start the dust ball with an infinite radius, which corresponds to (\ref{r0B9}) with $-$ signature.
The dust ball continues to collapse and when the radius coincides with the radius of the Schwarzschild black hole, there appears the horizon, which corresponds
to (\ref{r0B8}) with $-$ signature.
For the appearance of the horizon, it takes infinite time for the observer at an infinitely far place.
The time inside the horizon is disconnected with the time outside the horizon.
After the radius of the dust ball coincides with the radius, which corresponds to $+$ signature in (\ref{r0B8}), the dust ball continues to collapse and becomes
a point as in (\ref{r0B5}) with $-$ signature.

\section{de Sitter-Schwarzschild spacetime and collapsing star}\label{ads}

In this section, by using the formulation in the last section, we rewrite the de Sitter-Schwarzschild spacetime in the static coordinates
in the form that the black hole is embedded in the acceleratedly expanding universe, which is asymptotically de Sitter spacetime.
After that, we propose a model describing the collapse of a star in the asymptotically expanding 
de Sitter spacetime.

In the case of the de Sitter spacetime, where $\frac{a'(\tau)}{a(\tau)}=H_0$ $\left(H_0:\mbox{constant}\right)$,
Eqs.~(\ref{G5B}) and (\ref{GBivBB}) have the following forms,
\begin{align}
\label{G5BdS}
0 =&\, - H_0 r + \left( -1 + {H_0}^2 r^2 \right)
\frac{\partial \tau\left(t,r\right)}{\partial r} \, , \\
\label{GBivBBdS}
 - \e^{2\nu (r,t)} =&\, \left( -1 + H_0^2 r^2 \right)
\left( \frac{\partial \tau\left(t,r\right)}{\partial t} \right)^2 \, , \nonumber \\
\e^{2\lambda (r,t)} =&\, 1 - 2H_0 r \frac{\partial \tau\left(t,r\right)}{\partial r}
+ \left( -1 + {H_0}^2 r^2 \right) \left( \frac{\partial \tau\left(t,r\right)}{\partial r} \right)^2 \, .
\end{align}
Eq.~(\ref{G5BdS}) can be easily integrated to find
\begin{align}
\label{G5BdS2}
\tau\left(t,r\right)= \frac{1}{2H_0} \ln \left( 1 - {H_0}^2 r^2 \right) + \tau_0 \left(t\right)\, .
\end{align}
Here $\tau_0 \left( t \right)$ is an arbitrary function of $t$.
Then Eq.~(\ref{GBivBBdS}) gives
\begin{align}
\label{GBivBBdS2}
 - \e^{2\nu (r,t)} =&\, \left( -1 + H_0^2 r^2 \right) \tau_0'(t)^2 \, , \nonumber \\
\e^{2\lambda (r,t)} =&\, 1 + 2H_0 r \frac{H_0 r}{1 - {H_0}^2 r^2}
+ \left( -1 + {H_0}^2 r^2 \right) \left( \frac{H_0 r}{1 - {H_0}^2 r^2} \right)^2
= \frac{1}{1 - {H_0}^2 r^2}\, .
\end{align}
Then if we choose $\tau_0(t)=t$, we obtain a standard metric of the de Sitter spacetime in the static coordinate,
\begin{align}
\label{GBivBBdS3}
\e^{2\nu (r,t)} = 1 - H_0^2 r^2 \, , \quad
\e^{2\lambda (r,t)} = \frac{1}{1 - {H_0}^2 r^2}\, .
\end{align}
Because $r=\rho a_0 \e^{H_0 \tau}$, by using (\ref{G5BdS2}) with $\tau_0(t)=t$, we find
\begin{align}
\label{tdS}
t = \tau - \frac{1}{2H_0} \ln \left( 1 - {H_0}^2 {a_0}^2 \rho^2 \e^{2H_0 \tau} \right) \, ,
\end{align}
which gives the change of the coordinates.

In the static coordinates, the Schwarzschild-de Sitter spacetime is given by,
\begin{align}
\label{SdS1}
\e^{2\nu (r,t)} = 1 - \frac{2M}{r} - H_0^2 r^2 \, , \quad
\e^{2\lambda (r,t)} = \frac{1}{1 - \frac{2M}{r} - {H_0}^2 r^2}\, .
\end{align}
If we change the coordinates by
\begin{align}
\label{chvr}
r=\rho a_0 \e^{H_0 \tau}\, , \quad
t = \tau { -} \frac{1}{2H_0} \ln \left( 1 - \frac{2M}{a_0 \rho \e^{H_0\tau}}- {H_0}^2 {a_0}^2 \rho^2 \e^{2H_0 \tau} \right) \, ,
\end{align}
the line element is expressed as,
\begin{align}
\label{SdS2}
ds^2=&\, - \frac{\left( 1 - \frac{3M}{a_0 \rho \e^{H_0\tau}} + H_0 \rho a_0 \e^{H_0 \tau} \right)
\left( 1 - \frac{3M}{a_0 \rho \e^{H_0\tau}} - H_0 \rho a_0 \e^{H_0 \tau} \right)
}{1 - \frac{2M}{a_0 \rho \e^{H_0\tau}}- {H_0}^2 {a_0}^2 \rho^2 \e^{2H_0 \tau}} d\tau^2 \nonumber \\
&\, + \frac{\left( 1 - \frac{3M}{a_0 \rho \e^{H_0\tau}}\right) \frac{2M}{H_0 a_0 \rho^2 \e^{H_0\tau}}
+ 6M H_0 a_0 \e^{H_0 \tau}
}{1 - \frac{2M}{a_0 \rho \e^{H_0\tau}}- {H_0}^2 {a_0}^2 \rho^2 \e^{2H_0 \tau}} d\rho d\tau \nonumber \\
& + \frac{\frac{1}{4{H_0}^2} \left( \frac{2M}{a_0 \rho^2 \e^{H_0\tau}} - 2 {H_0}^2 {a_0}^2 \rho \e^{2H_0 \tau} + 2 H_0 a_0 \e^{H_0 \tau} \right)
\left( - \frac{2M}{a_0 \rho^2 \e^{H_0\tau}} + 2 {H_0}^2 {a_0}^2 \rho \e^{2H_0 \tau} + 2 H_0 a_0 \e^{H_0 \tau} \right)
}{1 - \frac{2M}{a_0 \rho \e^{H_0\tau}}- {H_0}^2 {a_0}^2 \rho^2 \e^{2H_0 \tau}} d\rho^2 \nonumber \\
&\, + {a_0}^2 \rho^2 \e^{2H_0} \left( d\vartheta^2 + \sin^2\vartheta d\varphi^2 \right) \, ,
\end{align}
which could describe the black hole in the asymptotically expanding de Sitter spacetime.
The change of the coordinate in (\ref{chvr}) is never unique but we just need to consider the change of the coordinate which becomes asymptotically to
coincide with (\ref{tdS}), that is, $t = \tau - \frac{1}{2H_0} \ln \left( 1 - {H_0}^2 {a_0}^2 \rho^2 \e^{2H_0 \tau} \right)$ with
$r=\rho a_0 \e^{H_0 \tau}$ for large $\rho$.

We now consider the model where a black hole is generated from the expanding universe.
Eq.~(\ref{tdS}) tells $t\to \tau$ when $\tau \to -\infty$ and $t\to + \infty$ when $1> {H_0}^2 {a_0}^2 \rho^2 \e^{2H_0 \tau} \to 1$, which correspond
to the cosmological horizon in the de Sitter spacetime.
An example is,
\begin{align}
\label{SdSdyn}
\e^{2\nu (r,t)} = \e^{- 2\lambda (r,t)} =1 - \frac{2M}{\sqrt{r^2+ {r_0}^2 \e^{- \frac{2t}{t_0}}}} - H_0^2 r^2 \, .
\end{align}
When $t\to -\infty$, the geometry becomes the de Sitter spacetime in (\ref{GBivBBdS3})
and when $t\to +\infty$, the geometry goes to the Schwarzschild-de Sitter spacetime in (\ref{SdS1}).
The black hole horizon, where $\e^{2\nu (r,t)} = \e^{- 2\lambda (r,t)} = 0$ appears when
\begin{align}
\label{SdSdyn1}
t = t_0 \ln \frac{2M}{r_0}\, .
\end{align}
and a black hole is formed.
We should note that the time of the black hole formation can be shifted by the redefinition of the time, for example, especially $t=0$, which corresponds to $r_0=2M$. 
The spacetime in (\ref{SdSdyn}) is, of course, different from the Oppenheimer-Snyder spacetime in the last subsection
because although the spacetime outside the collapsing dust in the Oppenheimer-Snyder model is the flat Minkowski spacetime.
On the other hand, in the model (\ref{SdSdyn}), the spacetime is asymptotically de Sitter spacetime, which can be written
in the form of an exponentially expanding universe. 
Then this model may describe the formation of the primordial black hole during the early stages of the universe. 
The Oppenheimer-Snyder model is, of course, realised by the dust but it has not been clarified how the spacetime given by (\ref{SdSdyn}) can be realised
by any physical process.
In the next section, we propose a model of the EGB theory including two scalar fields which realises the spacetime in (\ref{SdSdyn}).

\section{Ghost-free of EGB theory Combined With Two Scalar Fields}\label{SecII}

In Ref.~\cite{Nojiri:2020blr}, it has been shown that any spherically symmetric spacetime, even if the spacetime is
dynamical, can be realised by using Einstein's gravity coupled with two scalar fields.
However, the model described in~\cite{Nojiri:2020blr} includes unphysical degrees of freedom.
In this study, we construct models without ghosts by introducing constraints provided by the Lagrange multiplier fields, similar to that in
mimetic gravity~\cite{Chamseddine:2013kea}.
Furthermore, in order to consider the change in the speed at which the GW propagates, we also include the coupling of the two scalar
fields with the GB topological invariant as in Refs.~\cite{Nojiri:2005vv, Nojiri:2006je, Elizalde:2023rds}.
We should also note that if the spherically symmetric spacetime is asymptotically flat or asymptotically de Sitter spacetime,
any solution of the Einstein equation is also a solution in this model even if we include matter.

We start with the EGB theory coupled with the scalar fields $\phi$ and $\chi$, whose action is given by,
\begin{align}
\label{I8}
S_{\phi\chi} = \int d^4 x \sqrt{-g} & \left\{ \frac{R}{2\kappa^2}
 - \frac{1}{2} A (\phi,\chi) \partial_\mu \phi \partial^\mu \phi
 - B (\phi,\chi) \partial_\mu \phi \partial^\mu \chi \right. \nonumber \\
& \left. \qquad - \frac{1}{2} C (\phi,\chi) \partial_\mu \chi \partial^\mu \chi
 - V (\phi,\chi) - \xi(\phi, \chi) \mathcal{G} + \mathcal{L}_\mathrm{matter} \right\}\, .
\end{align}
In this action, $V(\phi,\chi)$ represents the potential, while $\xi(\phi,\chi)$ is also a function of $\phi$ and $\chi$.
In Eq.~(\ref{I8}), $\mathcal{G}$ refers to the GB invariant, which is defined by
\begin{align}
\label{eq:GB}
\mathcal{G} = R^2-4R_{\alpha \beta}R^{\alpha \beta}+R_{\alpha \beta \rho \sigma}R^{\alpha \beta \rho \sigma}\, ,
\end{align}
which is a topological density and becomes a total derivative in four dimensions that gives the Euler number.

Upon varying the action (\ref{I8}) with respect to the metric $g_{\mu\nu}$, we obtain,
\begin{align}
\label{gb4bD4}
0=&\, I_{\mu \nu} \nonumber \\
\equiv &\, \frac{1}{2\kappa^2}\left(- R_{\mu\nu} + \frac{1}{2} g_{\mu\nu} R\right) \nonumber \\
&\, + \frac{1}{2} g_{\mu\nu} \left\{
 - \frac{1}{2} A (\phi,\chi) \partial_\rho \phi \partial^\rho \phi
 - B (\phi,\chi) \partial_\rho \phi \partial^\rho \chi
 - \frac{1}{2} C (\phi,\chi) \partial_\rho \chi \partial^\rho \chi - V (\phi,\chi)\right\} \nonumber \\
&\, + \frac{1}{2} \left\{ A (\phi,\chi) \partial_\mu \phi \partial_\nu \phi
+ B (\phi,\chi) \left( \partial_\mu \phi \partial_\nu \chi
+ \partial_\nu \phi \partial_\mu \chi \right)
+ C (\phi,\chi) \partial_\mu \chi \partial_\nu \chi \right\} \nonumber \\
&\, - 2 \left( \nabla_\mu \nabla_\nu \xi(\phi,\chi)\right)R
+ 2 g_{\mu\nu} \left( \nabla^2 \xi(\phi,\chi)\right)R
+ 4 \left( \nabla_\rho \nabla_\mu \xi(\phi,\chi)\right)R_\nu^{\ \rho}
+ 4 \left( \nabla_\rho \nabla_\nu \xi(\phi,\chi)\right)R_\mu^{\ \rho} \nonumber \\
&\, - 4 \left( \nabla^2 \xi(\phi,\chi) \right)R_{\mu\nu}
 - 4g_{\mu\nu} \left( \nabla_\rho \nabla_\sigma \xi(\phi,\chi) \right) R^{\rho\sigma}
+ 4 \left(\nabla^\rho \nabla^\sigma \xi(\phi,\chi) \right) R_{\mu\rho\nu\sigma}
+ \frac{1}{2} T_{\mathrm{matter}\, \mu\nu} \, ,
\end{align}
and the field equations of the scalar fields are derived upon varying the action~(\ref{I8}) with respect to $\phi$ and $\chi$, yielding,
\begin{align}
\label{I10}
0 =& \frac{1}{2} A_\phi \partial_\mu \phi \partial^\mu \phi
+ A \nabla^\mu \partial_\mu \phi + A_\chi \partial_\mu \phi \partial^\mu \chi
+ \left( B_\chi - \frac{1}{2} C_\phi \right)\partial_\mu \chi \partial^\mu \chi
+ B \nabla^\mu \partial_\mu \chi - V_\phi - \xi_\phi \mathcal{G} \, ,\nonumber \\
0 =& \left( - \frac{1}{2} A_\chi + B_\phi \right) \partial_\mu \phi \partial^\mu \phi
+ B \nabla^\mu \partial_\mu \phi
+ \frac{1}{2} C_\chi \partial_\mu \chi \partial^\mu \chi + C \nabla^\mu \partial_\mu \chi
+ C_\phi \partial_\mu \phi \partial^\mu \chi - V_\chi - \xi_\chi \mathcal{G} \, .
\end{align}
Here $A_\phi=\partial A(\phi,\chi)/\partial \phi$, etc. and in
Eq.~(\ref{gb4bD4}), $T_{\mathrm{matter}\, \mu\nu}$ represents the matter energy-momentum tensor.
As well known, the field equations in Eq.~(\ref{I10}) correspond to the Bianchi identities.
By using the Bianchi identity, we always obtain $\nabla^\mu \left( - R_{\mu\nu} + \frac{1}{2}g_{\mu\nu} R \right)=0$, etc.
Because the four-vectors $\left( \partial_\mu \phi \right)$ and $\left( \partial_\mu \chi \right)$ are linearly independent of each other,
we may separate the equation $\partial^\nu I_{\nu\mu}=0$ for $I_{\mu\nu}$ in (\ref{gb4bD4}) into the part proportional to $\partial_\mu \phi$
and that proportional to $\partial_\mu \chi$ and we obtain the field equations (\ref{I10}).

As proposed in \cite{Nojiri:2020blr}, we also make the following identifications,
\begin{align}
\label{TSBH1}
\phi=t\, , \quad \chi=r\, .
\end{align}
This assumption remains valid for the following reasons:
In the spherically symmetric solutions (\ref{GBiv0}) derived from theory (\ref{I8}), typically, both $\phi$ and $\chi$ vary
with respect to both coordinates $t$ and $r$.
Once a solution is obtained, we can ascertain how $\phi$ and $\chi$ vary w.r.t. $t$ and $r$, and then express $\phi$ and $\chi$ as particular functions $\phi(t,r)$, $\chi(t,r)$ of $t$ and $r$.
By substituting $t$ and $r$ with new scalar fields,
denoted as $\bar{\phi}$ and $\bar{\chi}$, we may write $\phi(t,r)\to \phi(\bar{\phi}, \bar{\chi})$ and $\chi(t,r) \to \chi(\bar{\phi}, \bar{\chi})$.
Then, we can equate the new fields with $t$ and $r$ as shown in (\ref{TSBH1}), where $\bar{\phi}$ is replaced with $\phi=t$ and $\bar{\chi}$ with $\chi= r$.
The substitution of variables $\left( \phi , \chi\right) \rightarrow \left( \bar{\phi} , \bar{\chi} \right)$ can then be integrated into the redefinitions of $A$, $B$, $C$,
and $V$ within the action (\ref{I8}). Thus, the assumption (\ref{TSBH1}) does not result in a loss of generality.

As we will see later, in the framework of the two-scalar--EGB theory in (\ref{I8}), we can
construct a model that realises any given spherically symmetric
geometry in (\ref{GBiv0}), which could be static or time-dependent
but $A$ or $C$ often become negative in the realised geometry,
which indicates that $\phi$ or $\chi$ becomes a ghost degree of freedom.
Classically, the ghost mode possesses negative kinetic energy and generates states with negative norms in quantum theory.
Hence, the presence of the ghost mode suggests that the model lacks physical consistency.
Now, we remove the ghost modes by imposing constraints on $\phi$ and $\chi$, similar to the mimetic constraint described in~\cite{Chamseddine:2013kea},
rendering the ghost modes non-dynamical.

To remove the ghost modes using the Lagrange multiplier fields $\lambda_\phi$ and $\lambda_\chi$, we supplement the action $S_{\mathrm{GR} \phi\chi}$ in (\ref{I8})
with the additional terms $S_{\mathrm{GR} \phi\chi} \to S_{\mathrm{GR} \phi\chi} + S_\lambda$,
\begin{align}
\label{lambda1}
S_\lambda = \int d^4 x \sqrt{-g} \left[ \lambda_\phi \left( \e^{-2\nu(t=\phi, r=\chi)} \partial_\mu \phi \partial^\mu \phi + 1 \right)
+ \lambda_\chi \left( \e^{-2\lambda(t=\phi, r=\chi)} \partial_\mu \chi \partial^\mu \chi - 1 \right) \right] \, .
\end{align}
The constraints are obtained by varying $S_\lambda$ with respect to $\lambda_\phi$ and $\lambda_\chi$,
\begin{align}
\label{lambda2}
0 = \e^{-2\nu(t=\phi, r=\chi)} \partial_\mu \phi \partial^\mu \phi + 1 \, , \quad
0 = \e^{-2\lambda(t=\phi, r=\chi)} \partial_\mu \chi \partial^\mu \chi - 1 \, .
\end{align}
Consistently, these solutions are described by (\ref{TSBH1}).

The constraints outlined in Eq.~(\ref{lambda2}) render the scalar fields $\phi$ and $\chi$ non-dynamical, meaning that deviations of $\phi$ and $\chi$
from the background given in (\ref{TSBH1}) do not propagate.
Indeed, by examining the perturbations from (\ref{TSBH1}),
\begin{align}
\label{pert1}
\phi=t + \delta \phi \, , \quad \chi=r + \delta \chi\, ,
\end{align}
we obtain by using Eq.~(\ref{lambda2}),
\begin{align}
\label{pert2}
\partial_t \delta \phi = \partial_r \delta \chi = 0\, .
\end{align}
Thus, if we enforce the initial condition $\delta\phi=0$, we discover that $\delta\phi=0$ holds throughout the entire spacetime.
On the other hand, when we set the boundary condition $\delta\chi\to 0$ as $r\to \infty$, we observe that $\delta\chi=0$ across the entire spacetime.
These observations indicate that $\phi$ and $\chi$ are non-dynamical or fixed degrees of freedom.

It is important to emphasise that even in the modified model with the addition of $S_\lambda$, $\lambda_\phi=\lambda_\chi=0$ remains a valid solution.
Hence, any solution to the equations in (\ref{gb4bD4}) and (\ref{I10}) obtained from the original action (\ref{I8}) also serves as a solution for the model
whose action is adjusted by $S_{\mathrm{GR} \phi\chi} \to S_{\mathrm{GR} \phi\chi} + S_\lambda$ as outlined in (\ref{lambda1}).
We now validate this statement.

By the modification $S_{\mathrm{GR} \phi\chi} \to S_{\mathrm{GR} \phi\chi} + S_\lambda$ in (\ref{lambda1}),
the equations in (\ref{gb4bD4}) and (\ref{I10}) are modified as follows,
\begin{align}
\label{gb4bD4mod}
0= &\, \frac{1}{2\kappa^2}\left(- R_{\mu\nu} + \frac{1}{2} g_{\mu\nu} R\right) \nonumber \\
&\, + \frac{1}{2} g_{\mu\nu} \left\{
 - \frac{1}{2} A (\phi,\chi) \partial_\rho \phi \partial^\rho \phi
 - B (\phi,\chi) \partial_\rho \phi \partial^\rho \chi
 - \frac{1}{2} C (\phi,\chi) \partial_\rho \chi \partial^\rho \chi - V (\phi,\chi)\right\} \nonumber \\
&\, + \frac{1}{2} \left\{ A (\phi,\chi) \partial_\mu \phi \partial_\nu \phi
+ B (\phi,\chi) \left( \partial_\mu \phi \partial_\nu \chi
+ \partial_\nu \phi \partial_\mu \chi \right)
+ C (\phi,\chi) \partial_\mu \chi \partial_\nu \chi \right\} \nonumber \\
&\, - 2 \left( \nabla_\mu \nabla_\nu \xi(\phi,\chi)\right)R
+ 2 g_{\mu\nu} \left( \nabla^2 \xi(\phi,\chi)\right)R
+ 4 \left( \nabla_\rho \nabla_\mu \xi(\phi,\chi)\right)R_\nu^{\ \rho}
+ 4 \left( \nabla_\rho \nabla_\nu \xi(\phi,\chi)\right)R_\mu^{\ \rho} \nonumber \\
&\, - 4 \left( \nabla^2 \xi(\phi,\chi) \right)R_{\mu\nu}
 - 4g_{\mu\nu} \left( \nabla_\rho \nabla_\sigma \xi(\phi,\chi) \right) R^{\rho\sigma}
+ 4 \left(\nabla^\rho \nabla^\sigma \xi(\phi,\chi) \right) R_{\mu\rho\nu\sigma} \nonumber \\
&\, + \frac{1}{2}g_{\mu\nu} \left\{ \lambda_\phi \left( \e^{-2\nu(r=\chi)} \partial_\rho \phi \partial^\rho \phi + 1 \right)
+ \lambda_\chi \left( \e^{-2\lambda(r=\chi)} \partial_\rho \chi \partial^\rho \chi - 1 \right) \right\} \nonumber \\
&\, - \lambda_\phi \e^{-2\nu(r=\chi)} \partial_\mu \phi \partial_\nu \phi
 - \lambda_\chi \e^{-2\lambda(r=\chi)} \partial_\mu \chi \partial_\nu \chi
+ \frac{1}{2} T_{\mathrm{matter}\, \mu\nu} \, , \\
\label{I10mod}
0 =& \frac{1}{2} A_\phi \partial_\mu \phi \partial^\mu \phi
+ A \nabla^\mu \partial_\mu \phi + A_\chi \partial_\mu \phi \partial^\mu \chi
+ \left( B_\chi - \frac{1}{2} C_\phi \right)\partial_\mu \chi \partial^\mu \chi
+ B \nabla^\mu \partial_\mu \chi - V_\phi - \xi_\phi \mathcal{G} \nonumber \\
&\, - 2 \nabla^\mu \left( \lambda_\phi \e^{-2\nu(r=\chi)} \partial_\mu \phi \right) \, ,\nonumber \\
0 =& \left( - \frac{1}{2} A_\chi + B_\phi \right) \partial_\mu \phi \partial^\mu \phi
+ B \nabla^\mu \partial_\mu \phi
+ \frac{1}{2} C_\chi \partial_\mu \chi \partial^\mu \chi + C \nabla^\mu \partial_\mu \chi
+ C_\phi \partial_\mu \phi \partial^\mu \chi - V_\chi - \xi_\chi \mathcal{G} \nonumber \\
&\, - 2 \nabla^\mu \left( \lambda_\chi \e^{-2\lambda(r=\chi)} \partial_\mu \chi \right) \, .
\end{align}
We will now explore the solution to the equations in (\ref{gb4bD4}) and (\ref{I10}) under the assumption of (\ref{GBiv0}) and (\ref{TSBH1}). Next, we examine the components $(t,t)$ and $(r,r)$ in (\ref{gb4bD4mod}), yielding:
\begin{align}
\label{lambdas}
0=\lambda_\phi = \lambda_\chi\, .
\end{align}
The remaining components in (\ref{gb4bD4mod}) are fulfilled trivially. Conversely, Eq.~(\ref{I10mod}) provides:
\begin{align}
\label{lambdas2}
0 = \nabla^\mu \left( \lambda_\phi \e^{-2\nu(r=\chi)} \partial_\mu \phi \right)
= \nabla^\mu \left( \lambda_\chi \e^{-2\lambda(r=\chi)} \partial_\mu \chi \right) \, .
\end{align}
Equation~(\ref{lambdas2}) is satisfied if Eq.~(\ref{lambdas}) is verified.
This indicates that the solution to Eqs.~(\ref{gb4bD4}) and (\ref{I10}) also satisfies Eqs.~(\ref{gb4bD4mod}) and (\ref{I10mod}),
which correspond to the action being altered as $S_{\mathrm{GR} \phi\chi} \to S_{\mathrm{GR} \phi\chi} + S_\lambda$ in (\ref{lambda1}), provided that $\lambda_\phi$
and $\lambda_\chi$ vanish (\ref{lambdas}).
However, it is important to acknowledge that for Eqs.~(\ref{gb4bD4mod}) and (\ref{I10mod}), there might exist a solution where $\lambda_\phi$ and $\lambda_\chi$
do not vanish.

We now consider constructing a model with a solution corresponding
to the geometry (\ref{GBiv0}) based on~\cite{Nashed:2021cfs, Nojiri:2023qgd}.

Metric~(\ref{GBiv0}) yields the following non-vanishing connections,
\begin{align}
\label{GBv}
&\Gamma^t_{tt}=\dot\nu \, , \quad \Gamma^r_{tt} = \e^{-2(\lambda - \nu)}\nu' \, , \quad \Gamma^t_{tr}=\Gamma^t_{rt}=\nu'\, , \quad
\Gamma^t_{rr} = \e^{2\lambda - 2\nu}\dot\lambda \, , \quad \Gamma^r_{tr} = \Gamma^r_{rt} = \dot\lambda \, , \nonumber \\
& \Gamma^r_{rr}=\lambda'\, , \Gamma^i_{jk} = \bar{\Gamma}^i_{jk}\, ,\quad \Gamma^r_{ij}=-\e^{-2\lambda}r \bar{g}_{ij} \, , \quad
\Gamma^i_{rj}=\Gamma^i_{jr}=\frac{1}{r} \, \delta^i_{\ j}\,,
\end{align}
where the metric $\bar{g}_{ij}$ is given by
$\sum_{i,j=1}^2 \bar{g}_{ij} dx^i dx^j = d\vartheta^2 + \sin^2\vartheta , d\varphi^2$,
with $\left(x^1=\vartheta,\, x^2=\varphi\right)$, and
$\bar{ \Gamma}^i_{jk}$ represents the metric connection of $\bar{g}_{ij}$.
The ``dot'' and ``prime'' symbols denote differentiation with respect to $t$ and $r$, respectively.

Then we obtain the following non-vanishing components of ${I_\mu}^\nu$:
\begin{align}
\label{scalr}
&tt\,\mbox{-component:}\nonumber\\
&{\frac{\left( 2\lambda' r+ \e^{2 \lambda } - 1 \right) \e^{-2\lambda +2 \nu}}{\kappa^2 r^2}} \nonumber\\
&= \frac{A}{2} + \frac{C \e^{-2 \lambda +2 \nu}}2 + V \e^{2\nu}
+ \frac{8}{r^2} \left\{ \left( - \xi'' + \xi'\lambda' \right) \e^{-2 \lambda + 2 \nu}
+ \left( \xi'' - 3 \xi' \lambda' \right) {\e^{2 \nu -4\lambda}} + \dot\xi \dot\lambda - \dot\xi \dot\lambda \e^{-2 \lambda} \right\} \, ,\nonumber\\
&tr,\, rt\, \mbox{-component:}\nonumber\\
&{\frac{2\dot\lambda} {\kappa^2r}}=B + 8 \left\{ \left( \dot\xi' - \dot\xi \nu' - 3 \xi' \dot\lambda \right) \e^{-2 \lambda} - \dot\xi' + \dot\xi \nu' + \xi' \dot\lambda \right\}
\,,\nonumber\\
&rr\mbox{-component:}\nonumber\\
&\frac{2 r\nu' -{ \e^{2 \lambda }}+1}{\kappa^2 r^2}\nonumber\\
&=\frac{A \e^{-2 \nu +2 \lambda}}{2} + \frac{C}{2} - V \e^{2 \lambda}
+ 8 \left\{ \left( \dot\xi \dot\nu - \ddot\xi \right) \e^{-2\nu + 2 \lambda} - \dot\xi \dot\nu \e^{-2 \nu} - 3 \xi' \nu' \e^{-2 \lambda} + \xi' \nu' + \ddot\xi \e^{-2\nu} \right\}
\, ,\nonumber\\
&\vartheta\vartheta,\,\varphi\varphi\, \mbox{-component}:\nonumber\\
&\frac{2}{{\kappa^2}}\left[ \left\{ \nu'' r + \left( \nu' -\lambda' \right) \left( r \nu' +1 \right) \right\} \e^{-2 \lambda}
+ r \left( \dot\nu \dot\lambda -\ddot\lambda - \dot\lambda^2 \right) \e^{-2\nu} \right] \nonumber\\
&= r \left( A \e^{-2 \nu} - 2 V - C \e^{-2 \lambda} \right) + 16 \left[ \left\{ \xi' \ddot\lambda + 2 \dot\lambda \dot\xi' - \ddot\xi \lambda'
 - \dot\lambda \left( \dot\lambda +\dot\nu \right) \xi' - \dot\xi \left( \dot\lambda\nu' - \lambda'\dot\nu \right) \right\} \e^{-2 \nu -2 \lambda} \right. \nonumber \\
& \qquad \left. - \left\{ \xi' \nu'' + \left( \xi'' + \xi' \nu' -3 \xi' \lambda' \right) \nu' \right\} {\e^{-4\,\lambda }} \right] \, .
\end{align}
By solving Eq.~(\ref{scalr}) with respect to $A$, $B$, $C$, and $V$, we obtain the following explicit forms,
\begin{align}
\label{scalars}
A(t,r)=&\, \frac{\e^{-4\lambda }}{\kappa^2 r^2}
\left[ \left\{ r^2 \frac{\partial^2 \nu }{\partial r^2} + \left( r - r^2 \frac{\partial \nu }{\partial r} \right) \frac{\partial \lambda }{\partial r}
 -1 + \left( \frac{\partial \nu }{\partial r} \right)^2 r^2
+ r \frac{\partial \nu }{\partial r} \right\} \e^{2\nu + 2\lambda}
\right.\nonumber\\
&\, \left. + \e^{2\nu +4\lambda } + \left\{ r^2 \frac{\partial \nu }{\partial t} \frac{\partial \lambda }{\partial t} - r^2 \frac{\partial^2\lambda }{ \partial t^2}
 - r^2 \left( \frac{\partial \lambda }{\partial t} \right)^2 \right\} \e^{4\lambda} \right] \nonumber\\
&\, + \frac{8 \e^{-4\lambda}}{r^2} \left[ \left( \frac{\partial^2\xi}{ \partial r^2} - \frac{\partial \xi}{\partial r} \frac{\partial \lambda }{\partial r} \right) \e^{2\nu + 2\lambda}
+ \left\{ - r \frac{\partial \xi }{\partial r} \frac{\partial^2\lambda }{\partial {t}^2}
 - 2 r \frac{\partial \lambda}{\partial t} \frac{\partial^2\xi}{\partial t\partial r}
+ r \frac{\partial^2\xi}{\partial t^2} \frac{\partial \lambda}{\partial r} \right. \right. \nonumber\\
&\, \left. + r \left(\frac{\partial \lambda}{\partial t}\right)^2 \frac{\partial \xi}{\partial r}
+ \left( r \frac{\partial \xi}{\partial t} \frac{\partial \nu}{\partial r} + \frac{\partial \xi }{\partial t}
+ r \frac{\partial \xi }{\partial r} \frac{\partial \nu}{\partial t} \right) \frac{\partial \lambda}{\partial t}
 - r \frac{\partial \xi}{\partial t} \frac{\partial \nu}{\partial t} \frac{\partial \lambda }{\partial r} \right\} \e^{2 \lambda}
\nonumber \\
&\, \left. + \left\{ \left( r \frac{ \partial \nu }{\partial r}-1 \right) \frac{\partial^2\xi }{\partial r^2}
+ r \frac{\partial \xi }{\partial r} \frac{\partial^2\nu }{\partial r^2}
+ \left\{ 3\left( 1-r \frac{\partial \nu}{\partial r} \right) \frac{\partial \lambda}{\partial r}
+ r \left( \frac{\partial \nu}{ \partial r} \right)^2 \right\} \frac{\partial \xi}{\partial r} \right\} \e^{2 \nu}
 - \frac{\partial \xi}{\partial t} \frac{\partial \lambda}{\partial t} \e^{4\lambda} \right] \,,\nonumber\\
B(t,r)=&\, \frac{2}{\kappa^2 r} \frac{\partial \lambda }{\partial t}
+ \frac{8}{r^2} \left[ \left( \frac{\partial \xi }{\partial t} \frac{\partial \nu }{\partial r} +3 \frac{\partial \xi}{\partial r} \frac{\partial \lambda }{\partial t}
 - \frac{\partial^2\xi}{\partial t\partial r} \right) \e^{ -2\lambda} + \left( - \frac{ \partial \xi }{\partial r} \frac{ \partial \lambda }{\partial t}
+ \frac{\partial^2\xi }{\partial t \partial r} - \frac{\partial \xi }{\partial t} \frac{\partial \nu }{\partial r} \right\} \right] \,,\nonumber\\
C(t,r)
=&\, - \frac{\e^{2\lambda}}{r^2 \kappa^2} \left[ \left\{ r^2 \frac{\partial^2 \nu}{\partial r^2}
+ r^2\left( \frac{\partial \nu}{\partial r} \right)^2+ \left( -r- r^2 \frac{\partial \lambda}{\partial r} \right) \frac{\partial \nu}{\partial r}
 - r \frac{\partial \lambda}{\partial r} -1 \right\} \e^{ -2 \lambda} \right. \nonumber\\
&\, \left. + \left\{ r^2 \frac{\partial \nu}{\partial t} \frac{\partial \lambda}{\partial t} - r^2 \frac{\partial^2\lambda}{\partial t^2}
 - r^2\left( \frac{\partial \lambda}{\partial t} \right)^2 \right\} \e^{-2 \nu }+1 \right] \nonumber\\
&\, -\frac{8\e^{2\lambda}}{r^2} \left[ \left\{ \left( 1+r \frac{\partial \lambda}{\partial r} \right) \frac{\partial^2\xi}{\partial t^2}
 - r \frac{\partial \xi}{\partial r} \frac{\partial^2\lambda}{\partial t^2} -2r \frac{\partial \lambda}{\partial t} \frac{\partial^2\xi}{\partial t\partial r}
+r \frac{\partial \xi}{\partial t}\frac{\partial \lambda}{\partial t} \frac{\partial \nu}{\partial r} \right.\right.
\nonumber\\
&\, \left. + r \frac{\partial \lambda}{\partial t} \left( \frac{\partial \lambda}{\partial t} + \frac{\partial \nu}{ \partial t} \right) \frac{\partial \xi}{\partial r}
 - \frac{\partial \xi}{\partial t} \frac{\partial \nu}{\partial t} \left( 1+r \frac{\partial \lambda}{\partial r} \right) \right\} \e^{-2\lambda -2\nu}
+ \frac{\partial \xi}{\partial r} \frac{\partial \nu}{\partial r} \e^{ -2 \lambda} \nonumber\\
&\, \left. + \left\{ r \frac{\partial \xi}{\partial r} \frac{\partial^2\nu}{\partial r^2}
+ \left\{ r \frac{\partial^2 \xi}{\partial r^2} + \frac{\partial \xi}{\partial r} \left( r \frac{\partial \nu}{\partial r} - 3 -3r \frac{\partial \lambda}{\partial r} \right) \right\}
\frac{\partial \nu}{\partial r}\right\} \e^{-4\lambda }
+ \left( - \frac{\partial^2\xi}{\partial t^2} + \frac{\partial \xi}{\partial t} {\frac{\partial \nu}{\partial t}} \right) \e^{-2 \nu } \right]\,,\nonumber\\
V(t,r)=&\, - \frac{1}{\kappa^2 r^2}
\left\{ \left( - r \frac{\partial \lambda}{\partial r} + 1 + r \frac{\partial \nu}{\partial r} \right) \e^{-2 \lambda} - 1 \right\} \nonumber \\
&\, - \frac{4}{r^2}
\left[ \left\{ - \frac{\partial^2\xi}{\partial r^2} - \left( -{\frac{\partial \lambda }{ \partial r}} +{\frac{\partial \nu }{\partial r} } \right) {\frac{\partial \xi }{\partial r}}
\right\} \e^{-2 \lambda}
+ \left\{\frac{\partial^2\xi}{\partial r^2} + 3 \frac{\partial \xi}{\partial r} \left( -{\frac{\partial \lambda }{\partial r}} +{\frac{\partial \nu }{\partial r}} \right) \right\} \e^{-4\lambda}
\right. \nonumber \\
&\, \left. + \left\{ - \frac{\partial^2\xi}{\partial t^2} + \frac{\partial \xi}{\partial t} \left( - \frac{\partial \lambda}{\partial t} + \frac{\partial \nu}{\partial t} \right) \right\}
\e^{-2 \lambda -2 \nu}
 - \left\{ -{ \frac{\partial^2\xi }{\partial {t}^2}}+ {\frac{\partial \xi }{\partial t}} \left( -{\frac{\partial \lambda }{\partial t}}
+ \frac{\partial \nu }{\partial t} \right) \right\} \e^{-2 \nu} \right]\,. \end{align}
By replacing $t$ and $r$ with $\phi$ and $\chi$ in (\ref{scalars}), we obtain a model,
which has a solution in (\ref{GBiv0}) for given $\nu$ and $\lambda$.

We now consider if there are other kinds of solutions.
We assume the spacetime given by (\ref{GBiv0}) is asymptotically flat or (anti-)de Sitter spacetime when $t\to \pm \infty$ or $r\to \infty$.
In such a model realising the geometry, the static or eternal flat or (anti-)de Sitter spacetime is also a solution automatically.
This can be found by considering the limit $t_0\to \pm\infty$ after shifting time coordinate $t\to t_0 +t$ or the limit $r_0\to \infty$ after shifting the radial coordinate $r\to r_0 + r$.
In this limit, the scalar field $\phi=t_0 +t$ also goes to $\pm \infty$ or $\chi=r_0 + r$ goes to $\infty$
(Note that we can always redefine the scalar field so that the scalar field becomes finite in the limit.
For example, we can define new scalar fields as $\Phi=\frac{1}{\phi}$ or $X=\frac{1}{\chi}$) and $A$, $B$, and $C$ vanish and $V$ also vanishes
or becomes a constant corresponding to the effective cosmological constant.
This tells that in the limit $t_0\to \pm\infty$ or $r_0\to \infty$, the scalar fields decouple from gravity and
the model in this paper reduces to the standard Einstein gravity.
Therefore the standard cosmological solutions or self-gravitating objects such as a planet, the Sun, various types of stars, etc.,
in Einstein's gravity are surely also solutions in this model.
We should also note that because the model does not explicitly depend on any coordinate, the object like the extremely
compact star in this paper may be created in any place.
This could also tell that there should be solutions where several numbers of stars exist as long as the distances between them are large enough
so that we can neglect the non-linear contributions.
Therefore although it is impossible to find out all the solutions in the model, we find that there should be rich kinds of solutions in this model.

Now we are going to give a specific form of $\xi(t,r)$ where it is non-trivial only near the forming black hole or origin and only in the period of the black hole formation.
Then $\xi$ must become constant for large $r$ and in the infinite past and future.
A function that satisfies the above constraints can take the following form,
\begin{align}
\label{exp}
\xi(t,r)=\frac{\xi_0 r^2}{r^2+{r_0}^2\e^{- \frac{2t}{t_0}}}\,,
\end{align}
where $\xi_0$ is a constant, $t_0$, and $r_0$ are dimensional constant quantities. Using Eq.~(\ref{exp}) and (\ref{SdSdyn}), in (\ref{scalars}), we obtain the explicit form of
$A(t,r)$, $B(t,r)$, $C(t,r)$, and $V(t,r)$.
Furthermore, by replacing $(t,r)$ in the obtained expressions with $(\phi, \chi)$, we obtain the model of the two-scalar model.

We may calculate the scalars $A(t,r)$, $B(t,r)$, $C(t,r)$, and $V(t,r)$ using Eqs.~(\ref{SdSdyn}) and (\ref{exp}) in Eq.~(\ref{scalars}) but we obtain very lengthy expressions.
To show the physical properties of the forms $A(t,r)$, $B(t,r)$, $C(t,r)$, and $V(t,r)$, it could be sufficient to indicate the asymptotic forms of these quantities as $r\to \infty$.
The asymptote of $A(t,r)$, $B(t,r)$, $C(t,r)$, and $V(t,r)$ take the following forms,
\begin{align}
A(t,r) = &\, -\frac{16\e^{- \frac{2t}{t_0}} \left(\kappa^2\xi_0 {t_0}^2{H_0}^6 + 2\kappa^2\xi_0{H_0}^4 \right) {r_0}^2}{{H_0}^2\kappa^2{t_0}^2 r^2}
 -\frac{\e^{- \frac{2t}{t_0}} \left(96 \kappa^2\xi_0M{t_0}^2{H_0}^6+5M {t_0}^2{H_0}^4 \right) {r_0}^2}{{H_0}^2\kappa^2{t_0}^2 r^3}\nonumber\\
&\, +\frac{16\e^{- \frac{2t}{t_0}} \left\{ 2{H_0}^4 \left(\kappa^2\xi_0 {t_0}^2{H_0}^2+4\kappa^2 \xi_0 \right) {r_0}^2\e^{- \frac{2t}{t_0}}
+ \kappa^2\xi_0{t_0}^2{H_0}^4 \right\}{r_0}^2}{{H_0}^2\kappa^2{t_0}^2 r^4}\nonumber\\
&\, +\frac{\e^{-\frac{2t}{t_0}} \left\{ {H_0}^4 \left( 816\kappa^2\xi_0M{t_0}^2{H_0}^2+\frac{21}{2}M{t_0}^2 \right) {r_0}^2\e^{- \frac{2t}{t_0}}+2M
 - 544\kappa^2\xi_0M{t_0}^2{H_0}^4- \left(32\kappa^2\xi_0+5{t_0}^2 \right) M{H_0}^2 \right\} {r_0}^2}{{H_0}^2\kappa^2{t_0}^2 r^5} \nonumber \\
&\, +\mathcal{O}\left(\frac{1}{r^6}\right)\,,\nonumber\\
B(t,r)=&\, -\frac{48 \xi_0 {r_0}^2\e^{- \frac{2t}{t_0}}{H_0}^2}{{t_0} r^3}+\frac{16\xi_0 {r_0}^2
\e^{-\frac{2t}{t_0}} \left( 10{H_0}^2{r_0}^2\e^{- \frac{2t}{t_0}}-1 \right)}{ {t_0}r^5}+\mathcal{O}\left(\frac{1}{r^6}\right)\,,\nonumber\\
C(t,r)=&\,-\frac{48\xi_0{r_0}^2\e^{- \frac{2t}{t_0}}{H_0}^2}{ r^4}+\frac{ \left(160\kappa^2\xi_0{r_0}^2\e^{- \frac{2t}{t_0}}{H_0}^4
+16 \kappa^2\xi_0{H_0}^2 \right) {r_0}^2\e^{-\frac{2t}{t_0}}}{{H_0}^2\kappa^2 r^6}+\frac{5M {r_0}^2\e^{- \frac{2t}{t_0}}}{{H_0}^2\kappa^2 r^7}
+\mathcal{O}\left(\frac{1}{r^8}\right)\,,\nonumber\\
V(t,r)=&\, \frac{3{H_0}^2}{\kappa^2} - \frac{24{r_0}^2\e^{- \frac{2t}{t_0}}{H_0}^4\xi_0}{r^2} \nonumber\\
&\, + \frac{ 144 {r_0}^4\xi_0 \e^{-\frac{4t}{t_0}} {H_0}^4{t_0}^2+8 \left( \xi_0{t_0}^2{H_0}^2-2\xi_0 \right) {r_0}^2\e^{-\frac{2t}{t_0}}
 -128{H_0}^4{r_0}^4\xi_0\e^{-\frac{4t}{t_0}}{t_0}^2 }{ {t_0}^2 r^4}+\frac{2M\left(1+24{H_0}^2\kappa^2\xi_0 \right) {r_0}^2\e^{- \frac{2t}{t_0}}}{\kappa^2 r^5} \nonumber \\
&\, +\mathcal{O}\left(\frac{1}{r^6}\right)\,,
\end{align}
and as $t \to \infty$ we obtain
\begin{align}
A(t,r) \approx &\, - \frac{{r_0}^2\e^{- \frac{2t}{t_0}}}{ r^9 \left(2M -r+ r^3{H_0}^2 \right){t_0}^2\kappa^2}
\left[ 32{H_0}^4\kappa^2\xi_0 r^{10}+5{H_0}^4{t_0}^2M r^9-32{H_0}^2\kappa^2\xi_0 r^8-10{H_0}^2 r^7{t_0}^2M \right. \nonumber\\
&\, + 20{H_0}^2 r^6 M^2{t_0}^2 +2M r^7 +32 r^5\kappa^2\xi_0M -64 r^4\kappa^2\xi_0 M^2 +1920 M^4\kappa^2\xi_0{t_0}^2 +20 r^3{t_0}^2 M^3 \nonumber\\
&\, -20 r^4 M^2{t_0}^2 + 5 r^5{t_0}^2M + 32{H_0}^2 r^7\kappa^2 \xi_0 M + 16{H_0}^2 r^6 \kappa^2\xi_0{t_0}^2 - 32{H_0}^4\kappa^2\xi_0{t_0}^2 r^8
 -208 r^3\kappa^2 \xi_0 {t_0}^2M \nonumber\\
&\, +1312\kappa^2\xi_0{t_0}^2 M^2 r^2 - 2752\kappa^2\xi_0{t_0}^2 M^3r + 16{H_0}^6 \kappa^2\xi_0{t_0}^2 r^{10} + 96{H_0}^6\kappa^2\xi_0{t_0}^2M r^9
 - 336{H_0}^4\kappa^2\xi_0{t_0}^2M r^7 \nonumber\\
&\, \left. + 864 {H_0}^4\kappa^2\xi_0 {t_0}^2 M^2 r^6 +448 {H_0}^2 \kappa^2\xi_0 {t_0}^2M r^5-2112 {H_0}^2\kappa^2\xi_0 {t_0}^2 M^2 r^4
+2304 {H_0}^2\kappa^2\xi_0 {t_0}^2 M^3 r^3 \right] \,,\nonumber\\
B(t,r) \approx &\, - \frac{2{r_0}^2\e^{- \frac{2t}{t_0}}}{ r^6{t_0} \left( 2M+{H_0}^2 r^3- r \right) \kappa^2} \nonumber \\
&\, \times \left[72 r^3M{H_0}^2\xi_0\kappa^2 -32rM\xi_0\kappa^2+ r^3M+24 r^6\xi_0{H_0}^4\kappa^2-16{H_0}^2 r^4\kappa^2\xi_0+48\xi_0 M^2\kappa^2 \right] \,,\nonumber\\
C(t,r) \approx &\, - \frac{{r_0}^2\e^{- \frac{2t}{t_0}}}{ \left( 2M+{H_0}^2 r^3 -r \right)^3 r^7\kappa^2{t_0}^2} \left[ 48{H_0}^8 r^{12}\kappa^2\xi_0{t_0}^2
 -160{H_0}^6 r^{10}\kappa^2\xi_0{t_0}^2-5 {H_0}^4M{t_0}^2 r^9 \right. \nonumber\\
&\, +288{H_0}^6M{t_0}^2 r^9\kappa^2\xi_0 + 176 r^8\kappa^2\xi_0{t_0}^2{H_0}^4 + 10M r^7{t_0}^2{H_0}^2 - 528M r^7{t_0}^2{H_0}^4\kappa^2\xi_0-2M r^7 \nonumber\\
&\, + 96M r^7\kappa^2\xi_0{H_0}^2 + 288{H_0}^4{t_0}^2 r^6 M^2\kappa^2\xi_0 - 20{H_0}^2{t_0}^2 r^6 M^2 - 64{H_0}^2{t_0}^2 r^6 \kappa^2\xi_0
 - 5M r^5{t_0}^2 \nonumber \\
&\, + 128M r^5{t_0}^2\kappa^2\xi_0{H_0}^2 - 96M r^5\kappa^2\xi_0 + 20 M^2 r^4{t_0}^2 + 384 M^2 r^4{t_0}^2\kappa^2\xi_0{H_0}^2 +192 M^2 r^4\kappa^2\xi_0 \nonumber\\
&\, -20 M^3{t_0}^2 r^3 - 768 M^3{t_0}^2 r^3\kappa^2\xi_0{H_0}^2 + 112M{t_0}^2 r^3\kappa^2\xi_0 - 736 M^2 r^2\kappa^2\xi_0{t_0}^2 + 1600 M^3\xi_0{t_0}^2\kappa^2r \nonumber\\
&\, \left.-1152 M^4\kappa^2\xi_0{t_0}^2 \right]
\,,\nonumber\\
V(t,r) \approx &\, \frac{3{H_0}^2}{\kappa^2} \,,
\end{align}
and as $t\to -\infty$ we obtain,
\begin{align}
A(t,r) \approx&\, -\frac{M\e^{\frac{t}{t_0}}}{\kappa^2{t_0}^2{r_0} \left( {H_0}^2 r^2-1 \right) r^2} \left[2 {H_0}^4{t_0}^2 r^4-4 {H_0}^2{t_0}^2 r^2- r^2+2 {t_0}^2 \right]
 \,,\nonumber\\
B(t,r) \approx&\, -\frac{2 M \e^{\frac{t}{t_0}}}{{r_0} r t_0 \left({H_0}^2 r^2-1 \right)\kappa^2}
\,,\nonumber\\
C(t,r) \approx&\, \frac{ M \e^{\frac{t}{t_0}} \left( 2 {H_0}^4{t_0}^2 r^4-4 {H_0}^2{t_0}^2 r^2- r^2+2 {t_0}^2 \right) }{{r_0} r^2{t_0}^2\kappa^2 \left( H_0^2 \,r^2-1 \right)^3 }
\,,\nonumber\\
V(t,r) \approx&\, \frac{3{H_0}^2}{\kappa^2}\, .
\end{align}
Therefore $A$, $B$, and $C$ vanish asymptotically and $V$ becomes a constant as expected from the geometry given by (\ref{SdSdyn}).
This tells that any solution of Einstein's gravity with a positive cosmological constant is surely also a solution of this model even if there exists matter.

\section{Propagation of Gravitational Wave}\label{SecIII}

Let us discuss the propagation of the GW now.
The requirement that the propagation speed of GW in the frame of the EGB theory coupled with a single scalar field matches the speed of light
in the FLRW metric spacetime has been provided in~\cite{Oikonomou:2020sij}.
Ref.~\cite{Nojiri:2023jtf} has given the condition for general spacetime.
We reobtain the condition in~\cite{Nojiri:2023jtf}, which also holds in the context of EGB theory combined with two scalar fields.

For the following general variation of the metric,
\begin{align}
\label{variation1}
g_{\mu\nu}\to g_{\mu\nu} + h_{\mu\nu}\,, 
\end{align}
the equation governing the propagation of the GW is derived in the following manner,
\begin{align}
\label{gb4bD4B0}
0=&\, \left[ \frac{1}{4\kappa^2} R + \frac{1}{2} \left\{
 - \frac{1}{2} A \partial_\rho \phi \partial^\rho \phi
 - B \partial_\rho \phi \partial^\rho \chi
 - \frac{1}{2} C \partial_\rho \chi \partial^\rho \chi - V \right\}
 - 4 \left( \nabla_{\rho} \nabla_\sigma \xi \right) R^{\rho\sigma} \right] h_{\mu\nu} \nonumber \\
&\, + \bigg[ \frac{1}{4} g_{\mu\nu} \left\{
 - A \partial^\tau \phi \partial^\eta \phi
 - B \left(\partial^\tau \phi \partial^\eta \chi + \partial^\eta \phi \partial^\tau \chi \right)
 - C \partial^\tau \chi \partial^\eta \chi \right\} \nonumber \\
&\, - 2 g_{\mu\nu} \left( \nabla^\tau \nabla^\eta \xi\right)R
 - 4 \left( \nabla^\tau \nabla_\mu \xi\right)R_\nu^{\ \eta} - 4 \left( \nabla^\tau \nabla_\nu \xi\right)R_\mu^{\ \eta}
+ 4 \left( \nabla^\tau \nabla^\eta \xi \right)R_{\mu\nu} \nonumber \\
&\, + 4g_{\mu\nu} \left( \nabla^\tau \nabla_\sigma \xi \right) R^{\eta\sigma}
+ 4g_{\mu\nu} \left( \nabla_\rho \nabla^\tau \xi \right) R^{\rho\eta}
 - 4 \left(\nabla^\tau \nabla^\sigma \xi \right) R_{\mu\ \, \nu\sigma}^{\ \, \eta}
 - 4 \left(\nabla^\rho \nabla^\tau \xi \right) R_{\mu\rho\nu}^{\ \ \ \ \eta}
\bigg] h_{\tau\eta} \nonumber \\
&\, + \frac{1}{2}\left\{ 2 \delta_\mu^{\ \eta} \delta_\nu^{\ \zeta} \left( \nabla_\kappa \xi \right)R
 - 2 g_{\mu\nu} g^{\eta\zeta} \left( \nabla_\kappa \xi \right)R
 - 4 \delta_\rho^{\ \eta} \delta_\mu^{\ \zeta} \left( \nabla_\kappa \xi \right)R_\nu^{\ \rho}
 - 4 \delta_\rho^{\ \eta} \delta_\nu^{\ \zeta} \left( \nabla_\kappa \xi \right)R_\mu^{\ \rho} \right. \nonumber \\
&\, \left. + 4 g^{\eta\zeta} \left( \nabla_\kappa \xi \right) R_{\mu\nu}
+ 4g_{\mu\nu} \delta_\rho^{\ \eta} \delta_\sigma^{\ \zeta} \left( \nabla_\kappa \xi \right) R^{\rho\sigma}
 - 4 g^{\rho\eta} g^{\sigma\zeta} \left( \nabla_\kappa \xi \right) R_{\mu\rho\nu\sigma}
\right\} g^{\kappa\lambda}\left( \nabla_\eta h_{\zeta\lambda} + \nabla_\zeta h_{\eta\lambda} - \nabla_\lambda h_{\eta\zeta} \right) \nonumber \\
&\, + \left\{ \frac{1}{4\kappa^2} g_{\mu\nu} - 2 \left( \nabla_\mu \nabla_\nu \xi\right) + 2 g_{\mu\nu} \left( \nabla^2\xi\right) \right\}
\left\{ -h_{\mu\nu} R^{\mu\nu} + \nabla^\mu \nabla^\nu h_{\mu\nu} - \nabla^2 \left(g^{\mu\nu}h_{\mu\nu}\right) \right\} \nonumber \\
&\, + \frac{1}{2}\left\{ \left( - \frac{1}{2\kappa^2} - 4 \nabla^2 \xi \right) \delta^\tau_{\ \mu} \delta^\eta_{\ \nu}
+ 4 \left( \nabla_\rho \nabla_\mu \xi\right) \delta^\eta_{\ \nu} g^{\rho\tau}
+ 4 \left( \nabla_\rho \nabla_\nu \xi\right) \delta^\tau_{\ \mu} g^{\rho\eta}
 - 4g_{\mu\nu} \nabla^\tau \nabla^\eta \xi \right\} \nonumber \\
&\, \qquad \times \left\{\nabla_\tau\nabla^\phi h_{\eta\phi}
+ \nabla_\eta \nabla^\phi h_{\tau\phi} - \nabla^2 h_{\tau\eta}
 - \nabla_\tau \nabla_\eta \left(g^{\phi\lambda}h_{\phi\lambda}\right)
 - 2R^{\lambda\ \phi}_{\ \eta\ \tau}h_{\lambda\phi}
+ R^\phi_{\ \tau}h_{\phi\eta} + R^\phi_{\ \tau}h_{\phi\eta} \right\} \nonumber \\
&\, + 2 \left(\nabla^\rho \nabla^\sigma \xi \right)
\left\{ \nabla_\nu \nabla_\rho h_{\sigma\mu}
 - \nabla_\nu \nabla_\mu h_{\sigma\rho}
 - \nabla_\sigma \nabla_\rho h_{\nu\mu}
 + \nabla_\sigma \nabla_\mu h_{\nu\rho}
+ h_{\mu\phi} R^\phi_{\ \rho\nu\sigma}
 - h_{\rho\phi} R^\phi_{\ \mu\nu\sigma} \right\} \nonumber \\
&\, + \frac{1}{2}\frac{\partial T_{\mathrm{matter}\,
\mu\nu}}{\partial g_{\tau\eta}}h_{\tau\eta} \, .
\end{align}
The equations in (\ref{gb4bD4B0}) are derived under the assumption that matter minimally couples with gravity.
We do not consider the scalar fields $\phi$ and $\chi$,
because the perturbation should vanish as we have observed in the last section (\ref{pert2}).

Next, we select a condition to fix the gauge degrees of freedom,
\begin{align}
\label{gfc}
0=\nabla^\mu h_{\mu\nu}\, .
\end{align}
As the present study is concerned with the massless spin-two mode, we further enforce the following constraint,
\begin{align}
\label{ce}
0=g^{\mu\nu} h_{\mu\nu} \, .
\end{align}
Applying the conditions in Eqs.~(\ref{gfc}) and (\ref{ce}), we can simplify Eq.~(\ref{gb4bD4B0}) as follows,
\begin{align}
\label{gb4bD4B}
0=&\, \left[ \frac{1}{4\kappa^2} R + \frac{1}{2} \left\{
 - \frac{1}{2} A \partial_\rho \phi \partial^\rho \phi
 - B \partial_\rho \phi \partial^\rho \chi
 - \frac{1}{2} C \partial_\rho \chi \partial^\rho \chi - V \right\}
 - 4 \left( \nabla_\rho \nabla_\sigma \xi \right) R^{\rho\sigma} \right] h_{\mu\nu} \nonumber \\
&\, + \bigg[ \frac{1}{4} g_{\mu\nu} \left\{
 - A \partial^\tau \phi \partial^\eta \phi
 - B \left(\partial^\tau \phi \partial^\eta \chi + \partial^\eta \phi \partial^\tau \chi \right)
 - C \partial^\tau \chi \partial^\eta \chi \right\} \nonumber \\
&\, - 2 g_{\mu\nu} \left( \nabla^\tau \nabla^\eta \xi\right)R
 - 4 \left( \nabla^\tau \nabla_\mu \xi\right)R_\nu^{\ \eta} - 4 \left( \nabla^\tau \nabla_\nu \xi\right)R_\mu^{\ \eta}
+ 4 \left( \nabla^\tau \nabla^\eta \xi \right)R_{\mu\nu} \nonumber \\
&\, + 4g_{\mu\nu} \left( \nabla^\tau \nabla_\sigma \xi \right) R^{\eta\sigma}
+ 4g_{\mu\nu} \left( \nabla_{\rho} \nabla^\tau \xi \right) R^{\rho\eta}
 - 4 \left(\nabla^\tau \nabla^\sigma \xi \right) R_{\mu\ \, \nu\sigma}^{\ \, \eta}
 - 4 \left(\nabla^\rho \nabla^\tau \xi \right) R_{\mu\rho\nu}^{\ \ \ \ \eta}
\bigg\} h_{\tau\eta} \nonumber \\
&\, + \frac{1}{2}\left\{ 2 \delta_\mu^{\ \eta} \delta_\nu^{\ \zeta} \left( \nabla_\kappa \xi \right)R
 - 4 \delta_\rho^{\ \eta} \delta_\mu^{\ \zeta} \left( \nabla_\kappa \xi \right)R_\nu^{\ \rho}
 - 4 \delta_\rho^{\ \eta} \delta_\nu^{\ \zeta} \left( \nabla_\kappa \xi \right)R_\mu^{\ \rho} \right. \nonumber \\
&\, \left. + 4g_{\mu\nu} \delta_\rho^{\ \eta} \delta_\sigma^{\ \zeta} \left( \nabla_\kappa \xi \right) R^{\rho\sigma}
 - 4 g^{\rho\eta} g^{\sigma\zeta} \left( \nabla_\kappa \xi \right) R_{\mu\rho\nu\sigma}
\right\} g^{\kappa\lambda}\left( \nabla_\eta h_{\zeta\lambda} + \nabla_\zeta h_{\eta\lambda} - \nabla_\lambda h_{\eta\zeta} \right) \nonumber \\
&\, - \left\{ \frac{1}{4\kappa^2} g_{\mu\nu} - 2 \left( \nabla_\mu \nabla_\nu \xi \right) + 2 g_{\mu\nu} \left( \nabla^2\xi \right) \right\}
R^{\mu\nu} h_{\mu\nu} \nonumber \\
&\, + \frac{1}{2}\left\{ \left( - \frac{1}{2\kappa^2} - 4 \nabla^2 \xi \right) \delta^\tau_{\ \mu} \delta^\eta_{\ \nu}
+ 4 \left( \nabla_\rho \nabla_\mu \xi \right) \delta^\eta_{\ \nu} g^{\rho\tau}
+ 4 \left( \nabla_\rho \nabla_\nu \xi \right) \delta^\tau_{\ \mu} g^{\rho\eta}
 - 4g_{\mu\nu} \nabla^\tau \nabla^\eta \xi \right\} \nonumber \\
&\, \qquad \times \left\{ - \nabla^2 h_{\tau\eta} - 2R^{\lambda\ \phi}_{\ \eta\ \tau}h_{\lambda\phi}
+ R^\phi_{\ \tau}h_{\phi\eta} + R^\phi_{\ \tau}h_{\phi\eta} \right\} \nonumber \\
&\, + 2 \left(\nabla^\rho \nabla^\sigma \xi \right)
\left\{ \nabla_\nu \nabla_\rho h_{\sigma\mu}
 - \nabla_\nu \nabla_\mu h_{\sigma\rho}
 - \nabla_\sigma \nabla_\rho h_{\nu\mu}
 + \nabla_\sigma \nabla_\mu h_{\nu\rho}
+ h_{\mu\phi} R^\phi_{\ \rho\nu\sigma}
 - h_{\rho\phi} R^\phi_{\ \mu\nu\sigma} \right\} \nonumber \\
&\, + \frac{1}{2}\frac{\partial T_{\mathrm{matter}\, \mu\nu}}{\partial g_{\tau\eta}}h_{\tau\eta} \, .
\end{align}
To investigate the propagation speed $c_\mathrm{GW}$ of the GW $h_{\mu\nu}$, we concentrate exclusively on the terms that include the second derivatives of $h_{\mu\nu}$.
\begin{align}
\label{second}
I_{\mu\nu} \equiv&\, I^{(1)}_{\mu\nu} + I^{(2)}_{\mu\nu} \, , \nonumber \\
I^{(1)}_{\mu\nu} \equiv&\, \frac{1}{2}\left\{ \left( - \frac{1}{2\kappa^2} - 4 \nabla^2 \xi \right) \delta^\tau_{\ \mu} \delta^\eta_{\ \nu}
+ 4 \left( \nabla_\rho \nabla_\mu \xi\right) \delta^\eta_{\ \nu} g^{\rho\tau}
+ 4 \left( \nabla_\rho \nabla_\nu \xi\right) \delta^\tau_{\ \mu} g^{\rho\eta}
 - 4g_{\mu\nu} \nabla^\tau \nabla^\eta \xi \right\} \nabla^2 h_{\tau\eta} \, , \nonumber \\
I^{(2)}_{\mu\nu} \equiv &\, 2 \left(\nabla^\rho \nabla^\sigma \xi \right)
\left\{ \nabla_\nu \nabla_\rho h_{\sigma\mu}
 - \nabla_\nu \nabla_\mu h_{\sigma\rho}
 - \nabla_\sigma \nabla_\rho h_{\nu\mu}
 + \nabla_\sigma \nabla_\mu h_{\nu\rho} \right\} \, .
\end{align}
Since we are assuming minimal coupling between matter and gravity, any matter contribution does not couple with derivatives of $h_{\mu\nu}$ and thus does not appear in $I_{\mu\nu}$.
In other words, the presence of matter does not affect the propagating speed of the GW.
It is worth noting that $I^{(1)}_{\mu\nu}$ does not alter the speed of the GW, which remains equal to the speed of light.
Generally, $I^{(2)}_{\mu\nu}$ can alter the speed of the GW from that of light.
If the metric $g_{\mu\nu}$ is proportional to $\nabla_\mu \nabla_\nu \xi$, i.e.,
\begin{align}
\label{condition}
\nabla_\mu \nabla_\nu \xi = \frac{1}{4}g_{\mu\nu} \nabla^2 \xi \, ,
\end{align}
$I^{(2)}_{\mu\nu}$ does not change the velocity of the GW from that of light.
The condition (\ref{condition}) is identical to that obtained in~\cite{Nojiri:2023jtf}
for the EGB theory coupled with a single scalar field.

Here we have not included the perturbations of the scalar fields
$\phi$ and $\chi$ because the perturbations can be put to vanish
due to the constraints (\ref{lambda2}).
In the case that there are no constraints, or in the case that only one scalar field is
coupled with the GB invariant, there are perturbations
of the scalar modes which are mixed with the scalar modes in the
fluctuation of the metric (\ref{variation1}).
As long as we consider the massless spin-two modes, which correspond to the
standard GW, the spin-two modes decouple
from the scalar mode at the leading order.
If we consider the second-order perturbation, the quadratic moment could appear by the
perturbation of the scalar mode. The quadratic moment plays the
role of the source of the GW.
Therefore, there could be a difference in the propagation of the GW
for the cases with and without the constraints (\ref{lambda2}).

In the metric (\ref{GBiv0}), only non-vanishing components of the connection are provided by (\ref{GBv}).
Employing the following definition of the Riemann tensor,
\begin{align}
\label{Riemann}
{R^\lambda}_{\ \mu\rho\nu} =\Gamma^\lambda_{\mu\nu,\rho} -\Gamma^\lambda_{\mu\rho,\nu} + \Gamma^\eta_{\mu\nu}\Gamma^\lambda_{\rho\eta}
 - \Gamma^\eta_{\mu\rho}\Gamma^\lambda_{\nu\eta} \,,
\end{align}
one finds,
\begin{align}
\label{curvatures}
R_{rtrt} = & - \e^{2\lambda} \left\{ \ddot\lambda + \left( \dot\lambda - \dot\nu \right) \dot\lambda \right\}
+ \e^{2\nu}\left\{ \nu'' + \left(\nu' - \lambda'\right)\nu' \right\} \, ,\nonumber \\
R_{titj} =& \, r\nu' \e^{-2\lambda + 2\nu} \bar{g}_{ij} \, ,\nonumber \\
R_{rirj} =& \, \lambda' r \bar{ g}_{ij} \, ,\quad {R_{tirj}= \dot\lambda r \bar{ g}_{ij} } \, , \quad
R_{ijkl} = \left( 1 - \e^{-2\lambda}\right) r^2 \left(\bar{g}_{ik} \bar{g}_{jl} - \bar{g}_{il} \bar{g}_{jk} \right)\, ,\nonumber \\
R_{tt} =& - \left\{ \ddot\lambda + \left( \dot\lambda - \dot\nu \right) \dot\lambda \right\}
+ \e^{-2\lambda + 2\nu} \left\{ \nu'' + \left(\nu' - \lambda'\right)\nu' + \frac{2\nu'}{r}\right\} \, ,\nonumber \\
R_{rr} =& \, \e^{2\lambda - 2\nu} \left\{ \ddot\lambda + \left( \dot\lambda - \dot\nu \right) \dot\lambda \right\}
 - \left\{ \nu'' + \left(\nu' - \lambda'\right)\nu' \right\}
+ \frac{2 \lambda'}{r} \, ,\quad
R_{tr} =R_{rt} = \frac{2\dot\lambda}{r} \, , \nonumber \\
R_{ij} =&\, \left\{ 1 + \left\{ - 1 - r \left(\nu' - \lambda' \right)\right\} \e^{-2\lambda}\right\} \bar{g}_{ij}\ , \nonumber \\
R=& \, 2 \e^{-2 \nu} \left\{ \ddot\lambda + \left( \dot\lambda -
\dot\nu \right) \dot\lambda \right\} + \e^{-2\lambda}\left\{ -
2\nu'' - 2\left(\nu' - \lambda'\right)\nu' - \frac{4\left(\nu' -
\lambda'\right)}{r} + \frac{2\e^{2\lambda} - 2}{r^2} \right\} \, ,
\end{align}
and
\begin{align}
\label{xis}
\nabla_t \nabla_t \xi =&\, \ddot\xi - \dot\nu \dot\xi - \e^{-2(\lambda - \nu)}\nu' \xi' \, , \quad
\nabla_r \nabla_r \xi = \xi'' - \e^{2\lambda - 2\nu} \dot\lambda \dot\xi - \lambda' \xi' \, , \quad
\nabla_t \nabla_r \xi = \nabla_r \nabla_t \xi = {\dot\xi}' - \nu' \dot\xi - \dot\lambda \xi' \, , \nonumber \\
\nabla_i \nabla_j \xi =&\, - \e^{-2\lambda} r {\bar g}_{ij} \xi'
\, , \quad \nabla^2 \xi = - \e^{-2\nu} \left( \ddot\xi - \left(
\dot\nu - \dot\lambda\right) \dot\xi \right) + \e^{-2\lambda}
\left( \xi'' + \left( \nu' - \lambda' - \frac{2}{r} \right) \xi'
\right) \, .
\end{align}

We now consider the GW which propagates in the radial direction,
\begin{align}
\label{hij} h_{ij} = \frac{\mathrm{Re} \left( \e^{-i\omega t + i k r} \right) h_{ij}^{(0)}}{r}
\quad \left( i,j =\vartheta, \varphi, \quad
\sum_i h_{\ \ i}^{(0)\ i}=0 \right)\, , \quad \mbox{other components}=0\, ,
\end{align}
and we also assume $k$ is large enough and therefore we keep the quadratic terms concerning $k$ and/or $\omega$ and neglect the terms which are
linear to $k$ or $\omega$.
Then Eq.~(\ref{second}) with (\ref{xis}) gives
\begin{align}
\label{dis1}
0=&\, \left[ - \frac{1}{4\kappa^2} - 2 \left\{ - \e^{-2\nu} \left( 2 \ddot\xi - \left( 2\dot\nu - \dot\lambda\right) \dot\xi \right) + \e^{-2\lambda}
\left( \xi'' + \left( 2 \nu' - \lambda' \right) \xi' \right) \right\} \right] \e^{-2\nu} \omega^2 \nonumber \\
&\, - \left[ - \frac{1}{4\kappa^2} - 2 \left\{ - \e^{-2\nu} \left( \ddot\xi - \left( \dot\nu - 2 \dot\lambda\right) \dot\xi \right) + \e^{-2\lambda}
\left( 2 \xi'' + \left( \nu' - 2 \lambda' \right) \xi' \right) \right\} \right] \e^{-2\lambda} k^2 \nonumber \\
&\, - 4 \left( {\dot\xi}' - \nu' \dot\xi - \dot\lambda \xi' \right) \e^{-2\nu - 2\lambda} k\omega \, .
\end{align}
If $\xi$ is small enough, we find,
\begin{align}
\label{dis2}
\frac{\omega}{k} = \e^{- \lambda +\nu } &\, \left[\pm\left\{ 1-4 \kappa^2 \left( -\ddot\xi + \dot\xi \left( \dot \lambda +\dot\nu \right) \right) \e^{ -2\nu}
 -4 \left\{ -\xi'' + \xi' \left(\lambda' +\nu' \right) \right\} \kappa^2 \e^{-2 \lambda} \right\} \right. \nonumber \\
&\, \left. - 8 \e^{-\lambda-\nu}\kappa^2 \left( \dot\xi \nu' + \xi' \dot\lambda -\dot\xi' \right)\right] \,.
\end{align}
Now, the speed of light is defined by $c=\e^{-\lambda+\nu}$.
Therefore, if the value of $\left\{ \left( \ddot\xi - \dot\xi \left( \dot \lambda +\dot\nu \right) \right) \e^{ -2\nu}
+ \left\{ \xi'' - \xi' \left(\lambda' +\nu' \right) \right\} \e^{-2\, \lambda} \right\}\pm2 \e^{-\lambda-\nu} \left( \dot\xi \nu' + \xi' \dot\lambda -\dot\xi' \right)$
is positive (negative), the speed at which the GW propagates is greater (smaller) than that of light. In Eq.~(\ref{dis2}),
the $+$ sign corresponds to the GW outwardly
propagating from the black hole, and the $-$ sign to that
inward-propagating into the black hole.
Therefore, the speed of GW propagating into the black hole, that is,
\begin{align}
\label{ds3}
v_\mathrm{in} \equiv \e^{- \lambda +\nu}\left[1-4 \kappa^2 \left\{ -\ddot\xi + \dot\xi \left( \dot \lambda +\dot\nu \right) \right\} \e^{ -2\nu}
 -4 \kappa^2 \left\{ -\xi'' + \xi' \left(\lambda' +\nu' \right) \right\} \e^{-2 \lambda} + 8 \e^{-\lambda-\nu}
\kappa^2 \left( \dot\xi \nu' + \xi' \dot\lambda -\dot\xi' \right)\right]\, ,
\end{align}
is different from the speed of GW propagating
outwards the black hole spacetime,
\begin{align}
\label{ds44}
v_\mathrm{out} \equiv \e^{- \lambda +\nu}\left[1-4 \kappa^2 \left\{ -\ddot\xi + \dot\xi \left( \dot \lambda +\dot\nu \right) \right\} \e^{ -2\nu}
 -4 \kappa^2 \left\{ -\xi'' + \xi' \left(\lambda' +\nu' \right) \right\} \e^{-2 \lambda} - 8 \e^{-\lambda-\nu}
\kappa^2 \left( \dot\xi \nu' + \xi' \dot\lambda -\dot\xi' \right)\right] \, .
\end{align}
This occurs when $\xi$ depends on the radial coordinate $r$ or scalar field $\chi$ and $\lambda$ or $\xi$ depends on the time $t$ or scalar field $\phi$.
If $ \dot\xi \nu' + \xi' \dot\lambda -\dot\xi'>0$, we find $v_\mathrm{in}> v_\mathrm{out}$ and if $ \dot\xi \nu' + \xi' \dot\lambda -\dot\xi'< 0$, $v_\mathrm{in}< v_\mathrm{out}$.
For example, we may consider the form of $\xi(t,r)$ given by Eq.~(\ref{exp}) and rewrite it in the form of $(\phi,\chi)$, i.e.,
\begin{align}
\label{ds4}
\xi(\phi,\chi)= 
\frac{\xi_0 \chi^2}{\chi^2+{r_0}^2 \e^{-\frac{2\phi}{t_0}}} \, .
\end{align}
Here $\xi_0$ and $t_0$ are positive constants. Under the choice (\ref{ds4}), we find $\xi\to 0$ when
$ t \to -\infty$ or $r\to 0$, which tells that
the propagation speed of the GW matches that of light in the distant past or at the centre of a black hole. We should also note
$\frac{\dot\xi}{\xi_0}>0$ if $\left|t\right|>0$, $\frac{\ddot\xi}{\xi_0}>0$ if $t<\frac{t_0}{2}$, and
$\frac{\ddot\xi}{\xi_0}<0$ if $t>\frac{t_0}{2}$. We further
find $\frac{\xi'}{\xi_0}$ always positive,
$\frac{\xi''}{\xi_0}<0$ if $t>-\frac{t_0}{2}$, $\frac{\dot{\xi'}}{\xi_0}<0$
if $t>\frac{t_0}{2}$, and $\frac{\dot{\xi'}}{\xi_0}>0$ if $t<\frac{t_0}{2}$.
On the other hand, we find
\begin{align}
\dot\lambda =&\, \frac{M {r_0}^2\e^{-\frac{2t}{t_0}}}{t_0\left( r^2+{r_0}^2\e^{-\frac{2t}{t_0}} \right)^\frac{3}{2}
\left( 1-{\frac{2M}{\sqrt { r^2+{r_0}^2\e^{-\frac{2t}{t_0}}}}}-{H_0}^2 r^2 \right)}>0\,, \nonumber \\
\lambda' =&\, \frac{ \left( {H_0}^2 \left( r^2+{r_0}^2\e^{- \frac{2t}{t_0}} \right)^\frac{3}{2}-M \right) r}{\left( r^2+{r_0}^2\e^{-\frac{2t}{t_0}} \right)^\frac{3}{2}
\left( 1-\frac{2M}{\sqrt { r^2+{r_0}^2\e^{- \frac{2t}{t_0}}}}-{H_0}^2 r^2 \right)}>0\,,
\end{align}
if we choose $M< {H_0}^2 \left( r^2+{r_0}^2\e^{- \frac{2t}{t_0}} \right)^\frac{3}{2}$.

It is complicated and not so useful if we discuss the general
choice of the parameters and general regions, but we may consider
some special simplified cases.
\begin{itemize}
\item First, we consider the region of the horizon $r_0\sim 2M$.
Then in Eqs.~(\ref{ds3}) and (\ref{ds44}), the terms including $\lambda$ and $\nu$ could be dominant but since
$\e^{2\nu}=\e^{-2\lambda} =0$, the terms including $\e^{2\nu}$ and $\e^{-2\lambda}$ become less dominant.
Therefore, the dominant terms could be given by,
\begin{align}
\label{ds5}
v_\mathrm{in} \sim v_\mathrm{out} \sim c \left[ 1 - 4\,\kappa^2 \dot\xi \left( \dot \lambda +\dot\nu \right) \right]\, .
\end{align}
It is crucial to highlight that the difference in speeds between inward-propagating and outward-propagating GWs becomes less notable. If $\xi_0 < 0$,
then we observe $\dot\xi (\dot \lambda + \dot\nu) < 0$, resulting in the propagating speed of the GW exceeding that of light.
Conversely, if $\xi_0 > 0$, the propagating speed decreases, falling below the speed of light.
\item We also consider the region where $r_0\gg 2M$.
Then we find,
\begin{align}
\label{ds6}
v_\mathrm{in} \sim v_\mathrm{out} \sim c \left[ 1 + 4\kappa^2 \xi'' \right]\, .
\end{align}
The difference between the speed of the inward-propagating GW and that of the outward-propagating one becomes
less dominant, again. If $\xi_0>0$, $\xi''<0$ in the region, and therefore the propagating of the speed of the GW decreases, becoming smaller than that of light again, and if $\xi_0<0$,
the propagating speed becomes larger than the speed of light.
\end{itemize}
In both of the above regions, the difference between the speed of the inward-propagating GW and that of the
outward-propagating one could be neglected.
Therefore, a difference could appear in the intermediate region.
We should also note that in both the above regions, if $\xi_0<0$, the propagating speed of the GW becomes larger than that of light
and if $\xi_2>0$, the propagating speed becomes smaller than the
speed of light.

Although the neutron star merging event of GW170817~\cite{TheLIGOScientific:2017qsa, Monitor:2017mdv, GBM:2017lvd} imposes strong constraints on 
the propagating speed of GW in the present universe, and therefore the Gauss-Bonnet coupling $\xi$, 
in the early universe, the propagating speed of GW might be drastically changed from that of light. 
If there exists such a change in the propagating speed, any foot stamp might be found in future observations. 

\section{Conclusion and discussion}

Within this study and to be able to create a PBH, we considered the gravitational collapse to the black hole in the acceleratingly expanding universe as in (\ref{SdSdyn})
in the frame of the EGB theory with two scalar fields and investigated the propagation of the GW.

In order to realise the spacetime of the gravitational collapse, we used the formulation of the ``reconstruction''.
One usually solves the equations for a given gravity model and finds its related structure of spacetime.
Here in this study, we considered the inverse procedure, that is, we find a model that realises the desired or given geometry.

The reconstructed models, however, often include ghosts.
The ghosts have the kinetic energy unbounded below in the classical theory and negative norm states in the quantum theory, which conflicts with
the so-called Copenhagen interpretation in the quantum theory.
Well-known ghosts in quantum field theory are the Fadeev-Popov ghosts in the gauge theories~\cite{Kugo:1979gm, Kugo:1977zq}.
The ghosts can be, however, eliminated by imposing constraints as in (\ref{lambda2}), which is similar to the mimetic constraint.
The obtained models have been applied to several kinds of spacetimes (for example, see~\cite{Nojiri:2023dvf}) as in this paper.
It has been shown that the standard cosmological solutions or self-gravitating objects such as a planet, the Sun, various types of stars, etc.,
in Einstein's gravity are also solutions in this model.

By choosing the GB coupling as in (\ref{exp}), we investigated the propagation of the high-frequency GW.
As discussed in~\cite{Oikonomou:2020sij, Nojiri:2023jtf}, the propagating speed changes due to the coupling during the period of the black hole formation.
In general, the propagating speed of the GW going into the black hole is different from that of the wave going out although the effects are
next-to-leading order.
By investigating the expression of the speeds, we have found the conditions that the propagating speed does not exceed the light speed and does not violate the causality.

It could be interesting to investigate the emission of the GW by the merger of two black holes and/or neutron stars
by using the formulation of the reconstruction.
We should note that we need four scalar fields because the spacetime has no symmetry~\cite{Nashed:2024jqw}.
We may consider the following action of $N$ scalar fields $\phi^{(a)}$ $\left( a = 1,2,\cdots,N \right)$,
\begin{align}
\label{st1}
S_\phi = \int d^4x \sqrt{-g} \left\{ \frac{1}{2} \sum_{a,b} A_{(ab)} \left( \phi^{(c)} \right) g^{\mu\nu} \partial_\mu \phi^{(a)} \partial_\nu \phi^{(b)} - V \left( \phi^{(c)} \right)
\right\} \, .
\end{align}
Note that there is an ambiguity in the redefinition of the scalar fields $\phi^{(a)} \to {\tilde\phi}^{(a)}\left( \phi^{(b)} \right)$, which changes $A_{(ab)}$ by
$A_{(ab)} \left( \phi^{(c)} \right) \to {\tilde A}_{(ab)} \left( {\tilde \phi}^{(c)} \right)\equiv
\sum_{d,f} \frac{\partial \phi^{(d)}}{\partial {\tilde \phi}^{(a)}} {\partial \phi^{(f)}}{\partial {\tilde \phi}^{(b)}} A_{(d f)} \left( \phi^{(h)}\left({\tilde \phi}^{(c)}\right) \right)$.
This change of the variables is an analogue of the coordinate transformation.
In general four-dimensional spacetime, the metric has ten components and four gauge degrees of freedom corresponding to the coordinate transformations.
Therefore we need at least four scalar fields so that the general spacetime can be realised in the framework of the scalar-tensor theory in (\ref{st1}),
as shown in~\cite{Nashed:2024jqw}.
When $N=4$ in (\ref{st1}), the functions $A_{(ab)} \left( \phi^{(c)} \right)$ has ten components and there is a potential $V \left( \phi^{(c)} \right)$.
Furthermore, there are four redundancies by the redefinition of the scalar fields.
This tells that the scalar-tensor theory (\ref{st1}) has one extra functional degree of freedom compared with the metric $g_{\mu\nu}$,
which allows us to choose $V=0$.
As a result, we obtain a non-linear $\sigma$ model, whose target space is a four-dimensional manifold.


\begin{thebibliography}{99}

\bibitem{Hawking:1971ei}
S.~Hawking,
Mon. Not. Roy. Astron. Soc. \textbf{152}, 75 (1971)
doi:10.1093/mnras/152.1.75

\bibitem{Zeldovich:1967lct}
Y.~B.~Zel'dovich and I.~D.~Novikov,
Sov. Astron. \textbf{10}, 602 (1967)

\bibitem{Carr:1974nx}
B.~J.~Carr and S.~W.~Hawking,
Mon. Not. Roy. Astron. Soc. \textbf{168}, 399-415 (1974)
doi:10.1093/mnras/168.2.399

\bibitem{Khlopov:2008qy}
M.~Y.~Khlopov,
Res. Astron. Astrophys. \textbf{10}, 495-528 (2010)
doi:10.1088/1674-4527/10/6/001
[arXiv:0801.0116 [astro-ph]].

\bibitem{Sasaki:2016jop}
M.~Sasaki, T.~Suyama, T.~Tanaka and S.~Yokoyama,
Phys. Rev. Lett. \textbf{117}, no.6, 061101 (2016)
[erratum: Phys. Rev. Lett. \textbf{121}, no.5, 059901 (2018)]
doi:10.1103/PhysRevLett.117.061101
[arXiv:1603.08338 [astro-ph.CO]].

\bibitem{Ali-Haimoud:2017rtz}
Y.~Ali-Ha\"\i{}moud, E.~D.~Kovetz and M.~Kamionkowski,
Phys. Rev. D \textbf{96}, no.12, 123523 (2017)
doi:10.1103/PhysRevD.96.123523
[arXiv:1709.06576 [astro-ph.CO]].

\bibitem{Harada:2016mhb}
T.~Harada, C.~M.~Yoo, K.~Kohri, K.~i.~Nakao and S.~Jhingan,
Astrophys. J. \textbf{833}, no.1, 61 (2016)
doi:10.3847/1538-4357/833/1/61
[arXiv:1609.01588 [astro-ph.CO]].

\bibitem{Harada:2017fjm}
T.~Harada, C.~M.~Yoo, K.~Kohri and K.~I.~Nakao,
Phys. Rev. D \textbf{96}, no.8, 083517 (2017)
[erratum: Phys. Rev. D \textbf{99}, no.6, 069904 (2019)]
doi:10.1103/PhysRevD.96.083517
[arXiv:1707.03595 [gr-qc]].

\bibitem{Khlopov:1980mg}
M.~Y.~Khlopov and A.~G.~Polnarev,
Phys. Lett. B \textbf{97}, 383-387 (1980)
doi:10.1016/0370-2693(80)90624-3

\bibitem{Khlopov:1985fch}
M.~Y.~Khlopov, B.~A.~Malomed, I.~B.~Zeldovich and Y.~B.~Zeldovich,
Mon. Not. Roy. Astron. Soc. \textbf{215}, no.4, 575-589 (1985)
doi:10.1093/mnras/215.4.575

\bibitem{Garcia-Bellido:1996mdl}
J.~Garcia-Bellido, A.~D.~Linde and D.~Wands,
Phys. Rev. D \textbf{54}, 6040-6058 (1996)
doi:10.1103/PhysRevD.54.6040
[arXiv:astro-ph/9605094 [astro-ph]].

\bibitem{Cotner:2019ykd}
E.~Cotner, A.~Kusenko, M.~Sasaki and V.~Takhistov,
JCAP \textbf{10}, 077 (2019)
doi:10.1088/1475-7516/2019/10/077
[arXiv:1907.10613 [astro-ph.CO]].

\bibitem{Martin:2019nuw}
J.~Martin, T.~Papanikolaou and V.~Vennin,
JCAP \textbf{01}, 024 (2020)
doi:10.1088/1475-7516/2020/01/024
[arXiv:1907.04236 [astro-ph.CO]].

\bibitem{Ezquiaga:2017fvi}
J.~M.~Ezquiaga, J.~Garcia-Bellido and E.~Ruiz Morales,
Phys. Lett. B \textbf{776}, 345-349 (2018)
doi:10.1016/j.physletb.2017.11.039
[arXiv:1705.04861 [astro-ph.CO]].

\bibitem{Carr:2018nkm}
B.~Carr, K.~Dimopoulos, C.~Owen and T.~Tenkanen,
Phys. Rev. D \textbf{97}, no.12, 123535 (2018)
doi:10.1103/PhysRevD.97.123535
[arXiv:1804.08639 [astro-ph.CO]].


\bibitem{Brout:1977ix}
R.~Brout, F.~Englert and E.~Gunzig,
Annals Phys. \textbf{115}, 78 (1978)
doi:10.1016/0003-4916(78)90176-8

\bibitem{Guth:1980zm}
A.~H.~Guth,
Phys. Rev. D \textbf{23}, 347-356 (1981)
doi:10.1103/PhysRevD.23.347

\bibitem{Starobinsky:1980te}
A.~A.~Starobinsky,
Phys. Lett. B \textbf{91}, 99-102 (1980)
doi:10.1016/0370-2693(80)90670-X

\bibitem{Linde:1981mu}
A.~D.~Linde,
Phys. Lett. B \textbf{108}, 389-393 (1982)
doi:10.1016/0370-2693(82)91219-9

\bibitem{Albrecht:1982wi}
A.~Albrecht and P.~J.~Steinhardt,
Phys. Rev. Lett. \textbf{48}, 1220-1223 (1982)
doi:10.1103/PhysRevLett.48.1220

\bibitem{Mukhanov:1981xt}
V.~F.~Mukhanov and G.~V.~Chibisov,
JETP Lett. \textbf{33}, 532-535 (1981)

\bibitem{Planck:2018jri}
Y.~Akrami \textit{et al.} [Planck],
Astron. Astrophys. \textbf{641}, A10 (2020)
doi:10.1051/0004-6361/201833887
[arXiv:1807.06211 [astro-ph.CO]].

\bibitem{BICEP:2021xfz}
P.~A.~R.~Ade \textit{et al.} [BICEP and Keck],
Phys. Rev. Lett. \textbf{127}, no.15, 151301 (2021)
doi:10.1103/PhysRevLett.127.151301
[arXiv:2110.00483 [astro-ph.CO]].

\bibitem{Caldwell:2022qsj}
R.~Caldwell, Y.~Cui, H.~K.~Guo, V.~Mandic, A.~Mariotti, J.~M.~No, M.~J.~Ramsey-Musolf, M.~Sakellariadou, K.~Sinha and L.~T.~Wang, \textit{et al.}
Gen. Rel. Grav. \textbf{54}, no.12, 156 (2022)
doi:10.1007/s10714-022-03027-x
[arXiv:2203.07972 [gr-qc]].

\bibitem{EPTA:2015qep}
L.~Lentati \textit{et al.} [EPTA],
Mon. Not. Roy. Astron. Soc. \textbf{453}, no.3, 2576-2598 (2015)
doi:10.1093/mnras/stv1538
[arXiv:1504.03692 [astro-ph.CO]].

\bibitem{Zhu:2014rta}
X.~J.~Zhu, G.~Hobbs, L.~Wen, W.~A.~Coles, J.~B.~Wang, R.~M.~Shannon, R.~N.~Manchester, M.~Bailes, N.~D.~R.~Bhat and S.~Burke-Spolaor, \textit{et al.}
Mon. Not. Roy. Astron. Soc. \textbf{444}, no.4, 3709-3720 (2014)
doi:10.1093/mnras/stu1717
[arXiv:1408.5129 [astro-ph.GA]].

\bibitem{Xu:2023wog}
H.~Xu, S.~Chen, Y.~Guo, J.~Jiang, B.~Wang, J.~Xu, Z.~Xue, R.~N.~Caballero, J.~Yuan and Y.~Xu, \textit{et al.}
Res. Astron. Astrophys. \textbf{23}, no.7, 075024 (2023)
doi:10.1088/1674-4527/acdfa5
[arXiv:2306.16216 [astro-ph.HE]].

\bibitem{Barausse:2020rsu}
E.~Barausse, E.~Berti, T.~Hertog, S.~A.~Hughes, P.~Jetzer, P.~Pani, T.~P.~Sotiriou, N.~Tamanini, H.~Witek and K.~Yagi, \textit{et al.}
Gen. Rel. Grav. \textbf{52}, no.8, 81 (2020)
doi:10.1007/s10714-020-02691-1
[arXiv:2001.09793 [gr-qc]].

\bibitem{LISA:2022kgy}
K.~G.~Arun \textit{et al.} [LISA],
Living Rev. Rel. \textbf{25}, no.1, 4 (2022)
doi:10.1007/s41114-022-00036-9
[arXiv:2205.01597 [gr-qc]].

\bibitem{Hu:2017mde}
W.~R.~Hu and Y.~L.~Wu,
Natl. Sci. Rev. \textbf{4}, no.5, 685-686 (2017)
doi:10.1093/nsr/nwx116

\bibitem{Ruan:2018tsw}
W.~H.~Ruan, Z.~K.~Guo, R.~G.~Cai and Y.~Z.~Zhang,
Int. J. Mod. Phys. A \textbf{35}, no.17, 2050075 (2020)
doi:10.1142/S0217751X2050075X
[arXiv:1807.09495 [gr-qc]].

\bibitem{TianQin:2015yph}
J.~Luo \textit{et al.} [TianQin],
Class. Quant. Grav. \textbf{33}, no.3, 035010 (2016)
doi:10.1088/0264-9381/33/3/035010
[arXiv:1512.02076 [astro-ph.IM]].

\bibitem{Boyle:2005se}
L.~A.~Boyle and P.~J.~Steinhardt,
Phys. Rev. D \textbf{77}, 063504 (2008)
doi:10.1103/PhysRevD.77.063504
[arXiv:astro-ph/0512014 [astro-ph]].

\bibitem{Kuroyanagi:2008ye}
S.~Kuroyanagi, T.~Chiba and N.~Sugiyama,
Phys. Rev. D \textbf{79}, 103501 (2009)
doi:10.1103/PhysRevD.79.103501
[arXiv:0804.3249 [astro-ph]].

\bibitem{Watanabe:2006qe}
Y.~Watanabe and E.~Komatsu,
Phys. Rev. D \textbf{73}, 123515 (2006)
doi:10.1103/PhysRevD.73.123515
[arXiv:astro-ph/0604176 [astro-ph]].

\bibitem{Saikawa:2018rcs}
K.~Saikawa and S.~Shirai,
JCAP \textbf{05}, 035 (2018)
doi:10.1088/1475-7516/2018/05/035
[arXiv:1803.01038 [hep-ph]].

\bibitem{Kuroyanagi:2008ye}
S.~Kuroyanagi, T.~Chiba and N.~Sugiyama,
Phys. Rev. D \textbf{79}, 103501 (2009)
doi:10.1103/PhysRevD.79.103501
[arXiv:0804.3249 [astro-ph]].

\bibitem{Parker:1974qw}
L.~Parker and S.~A.~Fulling,
Phys. Rev. D \textbf{9}, 341-354 (1974)
doi:10.1103/PhysRevD.9.341

\bibitem{Fulling:1974pu}
S.~A.~Fulling, L.~Parker and B.~L.~Hu,
Phys. Rev. D \textbf{10}, 3905-3924 (1974)
doi:10.1103/PhysRevD.10.3905

\bibitem{Durrer:2009ii}
R.~Durrer, G.~Marozzi and M.~Rinaldi,
Phys. Rev. D \textbf{80}, 065024 (2009)
doi:10.1103/PhysRevD.80.065024
[arXiv:0906.4772 [astro-ph.CO]].

\bibitem{Marozzi:2011da}
G.~Marozzi, M.~Rinaldi and R.~Durrer,
Phys. Rev. D \textbf{83}, 105017 (2011)
doi:10.1103/PhysRevD.83.105017
[arXiv:1102.2206 [astro-ph.CO]].

\bibitem{Markkanen:2017rvi}
T.~Markkanen,
JCAP \textbf{05}, 001 (2018)
doi:10.1088/1475-7516/2018/05/001
[arXiv:1712.02372 [hep-th]].

\bibitem{Wang:2015zfa}
D.~G.~Wang, Y.~Zhang and J.~W.~Chen,
Phys. Rev. D \textbf{94}, no.4, 044033 (2016)
doi:10.1103/PhysRevD.94.044033
[arXiv:1512.03134 [gr-qc]].

\bibitem{Zhang:2018dvc}
Y.~Zhang and B.~Wang,
JCAP \textbf{11}, 006 (2018)
doi:10.1088/1475-7516/2018/11/006
[arXiv:1805.12304 [gr-qc]].


\bibitem{Chapline:1975ojl}
G.~F.~Chapline,
Nature \textbf{253}, no.5489, 251-252 (1975)
doi:10.1038/253251a0

\bibitem{Ivanov:1994pa}
P.~Ivanov, P.~Naselsky and I.~Novikov,
Phys. Rev. D \textbf{50}, 7173-7178 (1994)
doi:10.1103/PhysRevD.50.7173

\bibitem{Bullock:1996at}
J.~S.~Bullock and J.~R.~Primack,
Phys. Rev. D \textbf{55}, 7423-7439 (1997)
doi:10.1103/PhysRevD.55.7423
[arXiv:astro-ph/9611106 [astro-ph]].

\bibitem{Kohri:2007qn}
K.~Kohri, D.~H.~Lyth and A.~Melchiorri,
JCAP \textbf{04}, 038 (2008)
doi:10.1088/1475-7516/2008/04/038
[arXiv:0711.5006 [hep-ph]].

\bibitem{Kawasaki:2016pql}
M.~Kawasaki, A.~Kusenko, Y.~Tada and T.~T.~Yanagida,
Phys. Rev. D \textbf{94}, no.8, 083523 (2016)
doi:10.1103/PhysRevD.94.083523
[arXiv:1606.07631 [astro-ph.CO]].
\bibitem{Garcia-Bellido:2017mdw}
J.~Garcia-Bellido and E.~Ruiz Morales,
Phys. Dark Univ. \textbf{18}, 47-54 (2017)
doi:10.1016/j.dark.2017.09.007
[arXiv:1702.03901 [astro-ph.CO]].

\bibitem{Germani:2017bcs}
C.~Germani and T.~Prokopec,
Phys. Dark Univ. \textbf{18}, 6-10 (2017)
doi:10.1016/j.dark.2017.09.001
[arXiv:1706.04226 [astro-ph.CO]].

\bibitem{Motohashi:2017kbs}
H.~Motohashi and W.~Hu,
Phys. Rev. D \textbf{96}, no.6, 063503 (2017)
doi:10.1103/PhysRevD.96.063503
[arXiv:1706.06784 [astro-ph.CO]].

\bibitem{Ballesteros:2017fsr}
G.~Ballesteros and M.~Taoso,
Phys. Rev. D \textbf{97}, no.2, 023501 (2018)
doi:10.1103/PhysRevD.97.023501
[arXiv:1709.05565 [hep-ph]].

\bibitem{Hertzberg:2017dkh}
M.~P.~Hertzberg and M.~Yamada,
Phys. Rev. D \textbf{97}, no.8, 083509 (2018)
doi:10.1103/PhysRevD.97.083509
[arXiv:1712.09750 [astro-ph.CO]].

\bibitem{Cai:2018tuh}
Y.~F.~Cai, X.~Tong, D.~G.~Wang and S.~F.~Yan,
Phys. Rev. Lett. \textbf{121}, no.8, 081306 (2018)
doi:10.1103/PhysRevLett.121.081306
[arXiv:1805.03639 [astro-ph.CO]].

\bibitem{Ballesteros:2018wlw}
G.~Ballesteros, J.~Beltran Jimenez and M.~Pieroni,
JCAP \textbf{06}, 016 (2019)
doi:10.1088/1475-7516/2019/06/016
[arXiv:1811.03065 [astro-ph.CO]].

\bibitem{Cai:2019bmk}
R.~G.~Cai, Z.~K.~Guo, J.~Liu, L.~Liu and X.~Y.~Yang,
JCAP \textbf{06}, 013 (2020)
doi:10.1088/1475-7516/2020/06/013
[arXiv:1912.10437 [astro-ph.CO]].

\bibitem{Linde:2012bt}
A.~Linde, S.~Mooij and E.~Pajer,
Phys. Rev. D \textbf{87}, no.10, 103506 (2013)
doi:10.1103/PhysRevD.87.103506
[arXiv:1212.1693 [hep-th]].

\bibitem{Bugaev:2013fya}
E.~Bugaev and P.~Klimai,
Phys. Rev. D \textbf{90}, no.10, 103501 (2014)
doi:10.1103/PhysRevD.90.103501
[arXiv:1312.7435 [astro-ph.CO]].

\bibitem{Domcke:2017fix}
V.~Domcke, F.~Muia, M.~Pieroni and L.~T.~Witkowski,
JCAP \textbf{07}, 048 (2017)
doi:10.1088/1475-7516/2017/07/048
[arXiv:1704.03464 [astro-ph.CO]].

\bibitem{Inomata:2017okj}
K.~Inomata, M.~Kawasaki, K.~Mukaida, Y.~Tada and T.~T.~Yanagida,
Phys. Rev. D \textbf{96}, no.4, 043504 (2017)
doi:10.1103/PhysRevD.96.043504
[arXiv:1701.02544 [astro-ph.CO]].

\bibitem{Lyth:2011kj}
D.~H.~Lyth,
[arXiv:1107.1681 [astro-ph.CO]].

\bibitem{Bugaev:2011wy}
E.~Bugaev and P.~Klimai,
Phys. Rev. D \textbf{85}, 103504 (2012)
doi:10.1103/PhysRevD.85.103504
[arXiv:1112.5601 [astro-ph.CO]].

\bibitem{Clesse:2015wea}
S.~Clesse and J.~Garc\'\i{}a-Bellido,
Phys. Rev. D \textbf{92}, no.2, 023524 (2015)
doi:10.1103/PhysRevD.92.023524
[arXiv:1501.07565 [astro-ph.CO]].

\bibitem{Kawasaki:2015ppx}
M.~Kawasaki and Y.~Tada,
JCAP \textbf{08}, 041 (2016)
doi:10.1088/1475-7516/2016/08/041
[arXiv:1512.03515 [astro-ph.CO]].

\bibitem{Oppenheimer:1939ue}
J.~R.~Oppenheimer and H.~Snyder,
Phys. Rev. \textbf{56}, 455-459 (1939)
doi:10.1103/PhysRev.56.455

\bibitem{Hawking:1970zqf}
S.~W.~Hawking and R.~Penrose,
Proc. Roy. Soc. Lond. A \textbf{314}, 529-548 (1970)
doi:10.1098/rspa.1970.0021

\bibitem{Hawking:1973uf}
S.~W.~Hawking and G.~F.~R.~Ellis,
Cambridge University Press, 2023,
ISBN 978-1-00-925316-1, 978-1-00-925315-4, 978-0-521-20016-5, 978-0-521-09906-6, 978-0-511-82630-6, 978-0-521-09906-6
doi:10.1017/9781009253161

\bibitem{Senovilla:2014gza}
J.~M.~M.~Senovilla and D.~Garfinkle,
Class. Quant. Grav. \textbf{32}, no.12, 124008 (2015)
doi:10.1088/0264-9381/32/12/124008
[arXiv:1410.5226 [gr-qc]].

\bibitem{Senovilla:1998oua}
J.~M.~M.~Senovilla,
Gen. Rel. Grav. \textbf{30}, 701 (1998)
doi:10.1023/A:1018801101244
[arXiv:1801.04912 [gr-qc]].

\bibitem{Malafarina:2017csn}
D.~Malafarina,
Universe \textbf{3}, no.2, 48 (2017)
doi:10.3390/universe3020048
[arXiv:1703.04138 [gr-qc]].

\bibitem{Joshi:2011rlc}
P.~S.~Joshi and D.~Malafarina,
Int. J. Mod. Phys. D \textbf{20}, 2641-2729 (2011)
doi:10.1142/S0218271811020792
[arXiv:1201.3660 [gr-qc]].

\bibitem{Mosani:2021hox}
K.~Mosani, D.~Dey, P.~S.~Joshi, G.~C.~Samanta, H.~Menon and V.~D.~Patel,
Class. Quant. Grav. \textbf{40}, no.14, 145015 (2023)
doi:10.1088/1361-6382/acd97a
[arXiv:2106.01773 [gr-qc]].

\bibitem{Shaikh:2018cul}
R.~Shaikh and P.~S.~Joshi,
Phys. Rev. D \textbf{98}, no.2, 024033 (2018)
doi:10.1103/PhysRevD.98.024033
[arXiv:1801.01993 [gr-qc]].

\bibitem{Abbas:2017tvh}
G.~Abbas and M.~Tahir,
Eur. Phys. J. C \textbf{77}, no.8, 537 (2017)
doi:10.1140/epjc/s10052-017-5114-0
[arXiv:1707.08472 [gr-qc]].

\bibitem{Chatterjee:2021zre}
A.~Chatterjee, A.~Ghosh and S.~C.~Jaryal,
Phys. Rev. D \textbf{106}, no.4, 044049 (2022)
doi:10.1103/PhysRevD.106.044049
[arXiv:2108.11680 [gr-qc]].

\bibitem{Manna:2019tql}
G.~Manna,
Eur. Phys. J. C \textbf{80}, no.9, 813 (2020)
doi:10.1140/epjc/s10052-020-8383-y
[arXiv:1911.11753 [gr-qc]].

\bibitem{Ziaie:2019jfl}
A.~H.~Ziaie, H.~Moradpour and S.~Ghaffari,
Phys. Lett. B \textbf{793}, 276-280 (2019)
doi:10.1016/j.physletb.2019.04.055
[arXiv:1901.03055 [gr-qc]].

\bibitem{Dadhich:2013bya}
N.~Dadhich, S.~G.~Ghosh and S.~Jhingan,
Phys. Rev. D \textbf{88}, 084024 (2013)
doi:10.1103/PhysRevD.88.084024
[arXiv:1308.4312 [gr-qc]].

\bibitem{Mkenyeleye:2014dwa}
M.~D.~Mkenyeleye, R.~Goswami and S.~D.~Maharaj,
Phys. Rev. D \textbf{90}, no.6, 064034 (2014)
doi:10.1103/PhysRevD.90.064034
[arXiv:1407.4309 [gr-qc]].

\bibitem{Harada:1999jf}
T.~Harada,
Pramana \textbf{63}, 741-754 (2004)
doi:10.1007/BF02705196
[arXiv:gr-qc/0407109 [gr-qc]].

\bibitem{Ashtekar:2014ife}
A.~Ashtekar and V.~Petkov,
Springer, 2014,
ISBN 978-3-642-41991-1, 978-3-642-41992-8
doi:10.1007/978-3-642-41992-8

\bibitem{Bamba:2011sm}
K.~Bamba, S.~Nojiri and S.~D.~Odintsov,
Phys. Lett. B \textbf{698}, 451-456 (2011)
doi:10.1016/j.physletb.2011.03.038
[arXiv:1101.2820 [gr-qc]].

\bibitem{Joshi:2008zz}
P.~S.~Joshi,
Cambridge University Press, 2012,
ISBN 978-1-107-40536-3, 978-0-521-87104-4, 978-0-511-37283-4
doi:10.1017/CBO9780511536274

\bibitem{Aslam:2007zz}
M.~J.~Aslam, F.~Hussain, A.~Qadir, Riazuddin and H.~Saleem,
doi:10.1142/6405

\bibitem{Torres:2015aga}
R.~Torres and F.~Fayos,
Phys. Lett. B \textbf{747}, 245-250 (2015)
doi:10.1016/j.physletb.2015.05.078
[arXiv:1503.07407 [gr-qc]].

\bibitem{Modesto:2010uh}
L.~Modesto, J.~W.~Moffat and P.~Nicolini,
Phys. Lett. B \textbf{695}, 397-400 (2011)
doi:10.1016/j.physletb.2010.11.046
[arXiv:1010.0680 [gr-qc]].

\bibitem{Modesto:2011kw}
L.~Modesto,
Phys. Rev. D \textbf{86}, 044005 (2012)
doi:10.1103/PhysRevD.86.044005
[arXiv:1107.2403 [hep-th]].

\bibitem{Husain:2008tc}
V.~Husain,
[arXiv:0808.0949 [gr-qc]].

\bibitem{Bambi:2013caa}
C.~Bambi, D.~Malafarina and L.~Modesto,
Phys. Rev. D \textbf{88}, 044009 (2013)
doi:10.1103/PhysRevD.88.044009
[arXiv:1305.4790 [gr-qc]].

\bibitem{Tippett:2011hz}
B.~K.~Tippett and V.~Husain,
Phys. Rev. D \textbf{84}, 104031 (2011)
doi:10.1103/PhysRevD.84.104031
[arXiv:1106.1118 [gr-qc]].

\bibitem{Bambi:2013gva}
C.~Bambi, D.~Malafarina and L.~Modesto,
Eur. Phys. J. C \textbf{74}, 2767 (2014)
doi:10.1140/epjc/s10052-014-2767-9
[arXiv:1306.1668 [gr-qc]].

\bibitem{Singh:2003au}
P.~Singh and A.~Toporensky,
Phys. Rev. D \textbf{69}, 104008 (2004)
doi:10.1103/PhysRevD.69.104008
[arXiv:gr-qc/0312110 [gr-qc]].

\bibitem{Bojowald:2008zzb}
M.~Bojowald,
Living Rev. Rel. \textbf{11}, 4 (2008)
doi:10.12942/lrr-2008-4

\bibitem{Guo:2015eho}
J.~Q.~Guo and P.~S.~Joshi,
Phys. Rev. D \textbf{94}, no.4, 044063 (2016)
doi:10.1103/PhysRevD.94.044063
[arXiv:1511.06161 [gr-qc]].

\bibitem{Fathi:2016lws}
M.~Fathi, S.~Jalalzadeh and P.~V.~Moniz,
Eur. Phys. J. C \textbf{76}, no.10, 527 (2016)
doi:10.1140/epjc/s10052-016-4373-5
[arXiv:1609.04488 [gr-qc]].

\bibitem{Hernandez:2018rxc}
J.~M.~Hern\'andez, M.~Bellini and C.~Moreno,
Phys. Dark Univ. \textbf{23}, 100251 (2019)
doi:10.1016/j.dark.2018.100251
[arXiv:1810.07385 [gr-qc]].

\bibitem{Kiefer:2019bxk}
C.~Kiefer, N.~Kwidzinski and D.~Piontek,
Eur. Phys. J. C \textbf{79}, no.8, 686 (2019)
doi:10.1140/epjc/s10052-019-7193-6
[arXiv:1903.04391 [gr-qc]].

\bibitem{Thebault:2022dmv}
K.~P.~Y.~Thebault,
Class. Quant. Grav. \textbf{40}, no.5, 055007 (2023)
doi:10.1088/1361-6382/acb752
[arXiv:2209.05905 [gr-qc]].

\bibitem{Casadio:2010fw}
R.~Casadio, S.~D.~H.~Hsu and B.~Mirza,
Phys. Lett. B \textbf{695}, 317-319 (2011)
doi:10.1016/j.physletb.2010.10.060
[arXiv:1008.2768 [gr-qc]].

\bibitem{Marto:2013soa}
J.~Marto, Y.~Tavakoli and P.~Vargas Moniz,
Int. J. Mod. Phys. D \textbf{24}, no.03, 1550025 (2015)
doi:10.1142/S021827181550025X
[arXiv:1308.4953 [gr-qc]].

\bibitem{Bojowald:2005qw}
M.~Bojowald, R.~Goswami, R.~Maartens and P.~Singh,
Phys. Rev. Lett. \textbf{95}, 091302 (2005)
doi:10.1103/PhysRevLett.95.091302
[arXiv:gr-qc/0503041 [gr-qc]].

\bibitem{Tavakoli:2013rna}
Y.~Tavakoli, J.~Marto and A.~Dapor,
Int. J. Mod. Phys. D \textbf{23}, no.7, 1450061 (2014)
doi:10.1142/S0218271814500618
[arXiv:1303.6157 [gr-qc]].

\bibitem{Hild:2010id}
S.~Hild, M.~Abernathy, F.~Acernese, P.~Amaro-Seoane, N.~Andersson,
K.~Arun, F.~Barone, B.~Barr, M.~Barsuglia and M.~Beker, \textit{et
al.}
Class. Quant. Grav. \textbf{28} (2011), 094013
doi:10.1088/0264-9381/28/9/094013 [arXiv:1012.0908 [gr-qc]].




\bibitem{Baker:2019nia}
J.~Baker, J.~Bellovary, P.~L.~Bender, E.~Berti, R.~Caldwell,
J.~Camp, J.~W.~Conklin, N.~Cornish, C.~Cutler and R.~DeRosa,
\textit{et al.}
[arXiv:1907.06482 [astro-ph.IM]].


\bibitem{Smith:2019wny}
T.~L.~Smith and R.~Caldwell,
Phys. Rev. D \textbf{100} (2019) no.10, 104055
doi:10.1103/PhysRevD.100.104055 [arXiv:1908.00546 [astro-ph.CO]].


\bibitem{Crowder:2005nr}
J.~Crowder and N.~J.~Cornish,
Phys. Rev. D \textbf{72} (2005), 083005
doi:10.1103/PhysRevD.72.083005 [arXiv:gr-qc/0506015 [gr-qc]].


\bibitem{Smith:2016jqs}
T.~L.~Smith and R.~Caldwell,
Phys. Rev. D \textbf{95} (2017) no.4, 044036
doi:10.1103/PhysRevD.95.044036 [arXiv:1609.05901 [gr-qc]].



\bibitem{Seto:2001qf}
N.~Seto, S.~Kawamura and T.~Nakamura,
Phys. Rev. Lett. \textbf{87} (2001), 221103
doi:10.1103/PhysRevLett.87.221103 [arXiv:astro-ph/0108011
[astro-ph]].


\bibitem{Kawamura:2020pcg}
S.~Kawamura, M.~Ando, N.~Seto, S.~Sato, M.~Musha, I.~Kawano,
J.~Yokoyama, T.~Tanaka, K.~Ioka and T.~Akutsu, \textit{et al.}
[arXiv:2006.13545 [gr-qc]].



\bibitem{Bull:2018lat}
A.~Weltman, P.~Bull, S.~Camera, K.~Kelley, H.~Padmanabhan,
J.~Pritchard, A.~Raccanelli, S.~Riemer-S\o{}rensen, L.~Shao and
S.~Andrianomena, \textit{et al.}
Publ. Astron. Soc. Austral. \textbf{37} (2020), e002
doi:10.1017/pasa.2019.42 [arXiv:1810.02680 [astro-ph.CO]].




\bibitem{LISACosmologyWorkingGroup:2022jok}
P.~Auclair \textit{et al.} [LISA Cosmology Working Group],
[arXiv:2204.05434 [astro-ph.CO]].




\bibitem{NANOGrav:2023gor}
G.~Agazie \textit{et al.} [NANOGrav],
Astrophys. J. Lett. \textbf{951} (2023) no.1, L8
doi:10.3847/2041-8213/acdac6 [arXiv:2306.16213 [astro-ph.HE]].



\bibitem{sunnynew}
S.~Vagnozzi,
JHEAp \textbf{39} (2023), 81-98 doi:10.1016/j.jheap.2023.07.001
[arXiv:2306.16912 [astro-ph.CO]].

\bibitem{CMB-S4:2016ple}
K.~N.~Abazajian \textit{et al.} [CMB-S4],
[arXiv:1610.02743 [astro-ph.CO]].



\bibitem{SimonsObservatory:2019qwx}
M.~H.~Abitbol \textit{et al.} [Simons Observatory],
Bull. Am. Astron. Soc. \textbf{51} (2019), 147 [arXiv:1907.08284
[astro-ph.IM]].


\bibitem{Nojiri:2010wj}
S.~Nojiri and S.~D.~Odintsov,
Phys. Rept. \textbf{505} (2011), 59-144
doi:10.1016/j.physrep.2011.04.001
[arXiv:1011.0544 [gr-qc]].

\bibitem{Nojiri:2017ncd}
S.~Nojiri, S.~D.~Odintsov and V.~K.~Oikonomou,
Phys. Rept. \textbf{692} (2017), 1-104
doi:10.1016/j.physrep.2017.06.001
[arXiv:1705.11098 [gr-qc]].

\bibitem{Hwang:2005hb}
 J.~c.~Hwang and H.~Noh,
 Phys.\ Rev.\ D {\bf 71} (2005) 063536
 doi:10.1103/PhysRevD.71.063536
 [gr-qc/0412126].

\bibitem{Nojiri:2006je}
 S.~Nojiri, S.~D.~Odintsov and M.~Sami,
 Phys.\ Rev.\ D {\bf 74} (2006) 046004
 doi:10.1103/PhysRevD.74.046004
 [hep-th/0605039].

\bibitem{Nojiri:2005vv}
 S.~Nojiri, S.~D.~Odintsov and M.~Sasaki,
 Phys.\ Rev.\ D {\bf 71} (2005) 123509
 doi:10.1103/PhysRevD.71.123509
 [hep-th/0504052].

\bibitem{Satoh:2007gn}
 M.~Satoh, S.~Kanno and J.~Soda,
 Phys.\ Rev.\ D {\bf 77} (2008) 023526
 doi:10.1103/PhysRevD.77.023526
 [arXiv:0706.3585 [astro-ph]].

\bibitem{Yi:2018gse}
 Z.~Yi, Y.~Gong and M.~Sabir,
 Phys.\ Rev.\ D {\bf 98} (2018) no.8, 083521
 doi:10.1103/PhysRevD.98.083521
 [arXiv:1804.09116 [gr-qc]].

\bibitem{Guo:2009uk}
 Z.~K.~Guo and D.~J.~Schwarz,
 Phys.\ Rev.\ D {\bf 80} (2009) 063523
 doi:10.1103/PhysRevD.80.063523
 [arXiv:0907.0427 [hep-th]].

\bibitem{Jiang:2013gza}
 P.~X.~Jiang, J.~W.~Hu and Z.~K.~Guo,
 Phys.\ Rev.\ D {\bf 88} (2013) 123508
 doi:10.1103/PhysRevD.88.123508
 [arXiv:1310.5579 [hep-th]].

\bibitem{Kanti:2015pda}
 P.~Kanti, R.~Gannouji and N.~Dadhich,
 Phys.\ Rev.\ D {\bf 92} (2015) no.4, 041302
 doi:10.1103/PhysRevD.92.041302
 [arXiv:1503.01579 [hep-th]].

\bibitem{vandeBruck:2017voa}
 C.~van de Bruck, K.~Dimopoulos, C.~Longden and C.~Owen,
 arXiv:1707.06839 [astro-ph.CO].

\bibitem{Kanti:1998jd}
 P.~Kanti, J.~Rizos and K.~Tamvakis,
 Phys.\ Rev.\ D {\bf 59} (1999) 083512
 doi:10.1103/PhysRevD.59.083512
 [gr-qc/9806085].

\bibitem{Pozdeeva:2020apf}
E.~O.~Pozdeeva, M.~R.~Gangopadhyay, M.~Sami, A.~V.~Toporensky and
S.~Y.~Vernov,
Phys. Rev. D \textbf{102} (2020) no.4, 043525
doi:10.1103/PhysRevD.102.043525 [arXiv:2006.08027 [gr-qc]].

\bibitem{Pozdeeva:2021iwc}
E.~O.~Pozdeeva and S.~Y.~Vernov,
Eur. Phys. J. C \textbf{81} (2021) no.7, 633
doi:10.1140/epjc/s10052-021-09435-8 [arXiv:2104.04995 [gr-qc]].

\bibitem{Koh:2014bka}
S.~Koh, B.~H.~Lee, W.~Lee and G.~Tumurtushaa,
Phys. Rev. D \textbf{90} (2014) no.6, 063527
doi:10.1103/PhysRevD.90.063527 [arXiv:1404.6096 [gr-qc]].

\bibitem{Bayarsaikhan:2020jww}
B.~Bayarsaikhan, S.~Koh, E.~Tsedenbaljir and G.~Tumurtushaa,
JCAP \textbf{11} (2020), 057 doi:10.1088/1475-7516/2020/11/057
[arXiv:2005.11171 [gr-qc]].

\bibitem{DeLaurentis:2015fea}
 M.~De Laurentis, M.~Paolella and S.~Capozziello,
 Phys.\ Rev.\ D {\bf 91} (2015) no.8, 083531
 doi:10.1103/PhysRevD.91.083531
 [arXiv:1503.04659 [gr-qc]].

\bibitem{Chervon:2019sey}
 Scalar Field Cosmology, S.~Chervon, I.~Fomin, V.~Yurov and
 A.~Yurov, World Scientific 2019, \\ doi:10.1142/11405

\bibitem{Nozari:2017rta}
 K.~Nozari and N.~Rashidi,
 Phys.\ Rev.\ D {\bf 95} (2017) no.12, 123518
 doi:10.1103/PhysRevD.95.123518
 [arXiv:1705.02617 [astro-ph.CO]].

\bibitem{Odintsov:2018zhw}
 S.~D.~Odintsov and V.~K.~Oikonomou,
 Phys.\ Rev.\ D {\bf 98} (2018) no.4, 044039
 doi:10.1103/PhysRevD.98.044039
 [arXiv:1808.05045 [gr-qc]].

 \bibitem{Kawai:1998ab}
 S.~Kawai, M.~a.~Sakagami and J.~Soda,
 Phys.\ Lett.\ B {\bf 437}, 284 (1998)
 doi:10.1016/S0370-2693(98)00925-3
 [gr-qc/9802033].

\bibitem{Yi:2018dhl}
 Z.~Yi and Y.~Gong,
 Universe {\bf 5} (2019) no.9, 200
 doi:10.3390/universe5090200
 [arXiv:1811.01625 [gr-qc]].

\bibitem{vandeBruck:2016xvt}
 C.~van de Bruck, K.~Dimopoulos and C.~Longden,
 Phys.\ Rev.\ D {\bf 94} (2016) no.2, 023506
 doi:10.1103/PhysRevD.94.023506
 [arXiv:1605.06350 [astro-ph.CO]].

\bibitem{Kleihaus:2019rbg}
B.~Kleihaus, J.~Kunz and P.~Kanti,
Phys. Lett. B \textbf{804} (2020), 135401
doi:10.1016/j.physletb.2020.135401 [arXiv:1910.02121 [gr-qc]].


\bibitem{Bakopoulos:2019tvc}
 A.~Bakopoulos, P.~Kanti and N.~Pappas,
 Phys.\ Rev.\ D {\bf 101} (2020) no.4, 044026
 doi:10.1103/PhysRevD.101.044026
 [arXiv:1910.14637 [hep-th]].

\bibitem{Maeda:2011zn}
 K.~i.~Maeda, N.~Ohta and R.~Wakebe,
 Eur.\ Phys.\ J.\ C {\bf 72} (2012) 1949
 doi:10.1140/epjc/s10052-012-1949-6
 [arXiv:1111.3251 [hep-th]].


\bibitem{Bakopoulos:2020dfg}
A.~Bakopoulos, P.~Kanti and N.~Pappas,
Phys. Rev. D \textbf{101} (2020) no.8, 084059
doi:10.1103/PhysRevD.101.084059 [arXiv:2003.02473 [hep-th]].

\bibitem{Ai:2020peo}
W.~Y.~Ai,
Commun. Theor. Phys. \textbf{72} (2020) no.9, 095402
doi:10.1088/1572-9494/aba242 [arXiv:2004.02858 [gr-qc]].

\bibitem{Odintsov:2020xji}
S.~D.~Odintsov, V.~K.~Oikonomou and F.~P.~Fronimos,
Annals Phys. \textbf{420} (2020), 168250
doi:10.1016/j.aop.2020.168250 [arXiv:2007.02309 [gr-qc]].

\bibitem{Oikonomou:2020sij}
V.~K.~Oikonomou and F.~P.~Fronimos,
Class. Quant. Grav. \textbf{38} (2021) no.3, 035013
doi:10.1088/1361-6382/abce47 [arXiv:2006.05512 [gr-qc]].

\bibitem{Odintsov:2020zkl}
S.~D.~Odintsov and V.~K.~Oikonomou,
Phys. Lett. B \textbf{805} (2020), 135437
doi:10.1016/j.physletb.2020.135437 [arXiv:2004.00479 [gr-qc]].

\bibitem{Odintsov:2020mkz}
S.~D.~Odintsov, V.~K.~Oikonomou, F.~P.~Fronimos and
S.~A.~Venikoudis,
Phys. Dark Univ. \textbf{30} (2020), 100718
doi:10.1016/j.dark.2020.100718 [arXiv:2009.06113 [gr-qc]].

\bibitem{Venikoudis:2021irr}
S.~A.~Venikoudis and F.~P.~Fronimos,
Eur. Phys. J. Plus \textbf{136} (2021) no.3, 308
doi:10.1140/epjp/s13360-021-01298-y [arXiv:2103.01875 [gr-qc]].

\bibitem{Kong:2021qiu}
S.~B.~Kong, H.~Abdusattar, Y.~Yin and Y.~P.~Hu,
[arXiv:2108.09411 [gr-qc]].

\bibitem{Easther:1996yd}
 R.~Easther and K.~i.~Maeda,
 Phys.\ Rev.\ D {\bf 54} (1996) 7252
 doi:10.1103/PhysRevD.54.7252
 [hep-th/9605173].

\bibitem{Antoniadis:1993jc}
 I.~Antoniadis, J.~Rizos and K.~Tamvakis,
 Nucl.\ Phys.\ B {\bf 415} (1994) 497
 doi:10.1016/0550-3213(94)90120-1
 [hep-th/9305025].

\bibitem{Antoniadis:1990uu}
I.~Antoniadis, C.~Bachas, J.~R.~Ellis and D.~V.~Nanopoulos,
Phys.\ Lett.\ B \textbf{257} (1991), 278-284
doi:10.1016/0370-2693(91)91893-Z

\bibitem{Kanti:1995vq}
P.~Kanti, N.~Mavromatos, J.~Rizos, K.~Tamvakis and E.~Winstanley,
Phys. Rev. D \textbf{54} (1996), 5049-5058
doi:10.1103/PhysRevD.54.5049 [arXiv:hep-th/9511071 [hep-th]].

\bibitem{Kanti:1997br}
P.~Kanti, N.~Mavromatos, J.~Rizos, K.~Tamvakis and E.~Winstanley,
Phys. Rev. D \textbf{57} (1998), 6255-6264
doi:10.1103/PhysRevD.57.6255 [arXiv:hep-th/9703192 [hep-th]].

\bibitem{Easson:2020mpq}
D.~A.~Easson, T.~Manton and A.~Svesko,
JCAP \textbf{10} (2020), 026 doi:10.1088/1475-7516/2020/10/026
[arXiv:2005.12292 [hep-th]].

\bibitem{Rashidi:2020wwg}
N.~Rashidi and K.~Nozari,
Astrophys. J. \textbf{890}, 58
doi:10.3847/1538-4357/ab6a10
[arXiv:2001.07012 [astro-ph.CO]].

\bibitem{Odintsov:2023aaw}
S.~D.~Odintsov, V.~K.~Oikonomou and F.~P.~Fronimos,
Phys. Rev. D \textbf{107} (2023), 08
doi:10.1103/PhysRevD.107.084007
[arXiv:2303.14594 [gr-qc]].

\bibitem{Odintsov:2023lbb}
S.~D.~Odintsov and T.~Paul,
Phys. Dark Univ. \textbf{42} (2023), 101263
doi:10.1016/j.dark.2023.101263
[arXiv:2305.19110 [gr-qc]].

\bibitem{Odintsov:2023weg}
S.~D.~Odintsov, V.~K.~Oikonomou, I.~Giannakoudi, F.~P.~Fronimos and E.~C.~Lymperiadou,
[arXiv:2307.16308 [gr-qc]].


\bibitem{Oikonomou:2022xoq}
V.~K.~Oikonomou,
Astropart. Phys. \textbf{141} (2022), 102718
doi:10.1016/j.astropartphys.2022.102718 [arXiv:2204.06304
[gr-qc]].

\bibitem{Nojiri:2023jtf}
S.~Nojiri and S.~D.~Odintsov,
[arXiv:2308.06731 [gr-qc]].

\bibitem{TerenteDiaz:2023kgc}
J.~J.~Terente D\'\i{}az, K.~Dimopoulos, M.~Kar\v{c}iauskas and
A.~Racioppi,
[arXiv:2310.08128 [gr-qc]].

\bibitem{Kawai:2023nqs}
S.~Kawai and J.~Kim,
[arXiv:2308.13272 [astro-ph.CO]].




\bibitem{Kawai:2021edk}
S.~Kawai and J.~Kim,
Phys. Rev. D \textbf{104} (2021) no.8, 083545
doi:10.1103/PhysRevD.104.083545 [arXiv:2108.01340 [astro-ph.CO]].



\bibitem{Kawai:2017kqt}
S.~Kawai and J.~Kim,
Phys. Lett. B \textbf{789} (2019), 145-149
doi:10.1016/j.physletb.2018.12.019 [arXiv:1702.07689 [hep-th]].


\bibitem{Choudhury:2023kam}
S.~Choudhury,
[arXiv:2307.03249 [astro-ph.CO]].








\bibitem{TheLIGOScientific:2017qsa}
B.~P.~Abbott \textit{et al.} [LIGO Scientific and Virgo],
Phys. Rev. Lett. \textbf{119} (2017) no.16, 161101
doi:10.1103/PhysRevLett.119.161101 [arXiv:1710.05832 [gr-qc]].

\bibitem{Monitor:2017mdv}
B.~P.~Abbott \textit{et al.} [LIGO Scientific, Virgo, Fermi-GBM
and INTEGRAL],
Astrophys. J. Lett. \textbf{848} (2017) no.2, L13
doi:10.3847/2041-8213/aa920c [arXiv:1710.05834 [astro-ph.HE]].

\bibitem{GBM:2017lvd}
 B.~P.~Abbott {\it et al.}
 ``Multi-messenger Observations of a Binary Neutron Star Merger,''
 Astrophys.\ J.\ {\bf 848} (2017) no.2, L12
 doi:10.3847/2041-8213/aa91c9
 [arXiv:1710.05833 [astro-ph.HE]].

\bibitem{Odintsov:2020sqy}
S.~D.~Odintsov, V.~K.~Oikonomou and F.~P.~Fronimos,
Nucl. Phys. B \textbf{958} (2020), 115135
doi:10.1016/j.nuclphysb.2020.115135 [arXiv:2003.13724 [gr-qc]].

\bibitem{Oikonomou:2021kql}
V.~K.~Oikonomou,
Class. Quant. Grav. \textbf{38} (2021) no.19, 195025
doi:10.1088/1361-6382/ac2168 [arXiv:2108.10460 [gr-qc]].

\bibitem{Oikonomou:2022ksx}
V.~K.~Oikonomou, P.~D.~Katzanis and I.~C.~Papadimitriou,
Class. Quant. Grav. \textbf{39} (2022) no.9, 095008
doi:10.1088/1361-6382/ac5eba [arXiv:2203.09867 [gr-qc]].



\bibitem{Nojiri:2020blr}
S.~Nojiri, S.~D.~Odintsov and V.~Faraoni,
Phys. Rev. D \textbf{103}, no.4, 044055 (2021)
doi:10.1103/PhysRevD.103.044055
[arXiv:2010.11790 [gr-qc]].

\bibitem{Chamseddine:2013kea}
A.~H.~Chamseddine and V.~Mukhanov,
JHEP \textbf{11}, 135 (2013)
doi:10.1007/JHEP11(2013)135
[arXiv:1308.5410 [astro-ph.CO]].


\bibitem{Nashed:2021cfs}
G.~G.~L.~Nashed and S.~Nojiri,
Eur. Phys. J. C \textbf{83} (2023) no.1, 68
doi:10.1140/epjc/s10052-022-11165-4
[arXiv:2112.13391 [gr-qc]].

\bibitem{Nojiri:2023qgd}
S.~Nojiri and G.~G.~L.~Nashed,
Phys. Rev. D \textbf{108} (2023) no.2, 024014
doi:10.1103/PhysRevD.108.024014
[arXiv:2306.14162 [gr-qc]].

\bibitem{Kugo:1979gm}
T.~Kugo and I.~Ojima,
Prog. Theor. Phys. Suppl. \textbf{66} (1979), 1-130
doi:10.1143/PTPS.66.1

\bibitem{Kugo:1977zq}
T.~Kugo and I.~Ojima,
Phys. Lett. B \textbf{73} (1978), 459-462
doi:10.1016/0370-2693(78)90765-7

\bibitem{Nojiri:2023dvf}
S.~Nojiri and G.~G.~L.~Nashed,
Phys. Rev. D \textbf{108} (2023) no.12, 124049
doi:10.1103/PhysRevD.108.124049
[arXiv:2309.12379 [hep-th]].

\bibitem{Elizalde:2023rds}
E.~Elizalde, S.~Nojiri, S.~D.~Odintsov and V.~K.~Oikonomou,
[arXiv:2312.02889 [gr-qc]].


\bibitem{Nashed:2024jqw}
G.~G.~L.~Nashed and S.~Nojiri,
[arXiv:2402.12860 [gr-qc]].


\end{thebibliography}
\end{document}